\documentclass{article}
\usepackage{amsmath,amssymb}
\usepackage{amsmath,amssymb,amsthm}
\usepackage{xypic}
\usepackage{authblk}
\usepackage{color}
\usepackage{graphicx}
\usepackage{float}
\usepackage{hyperref}
\usepackage{fancyhdr}
\hypersetup{colorlinks=true,linkcolor=black,citecolor=black}
 \usepackage[toc,page]{appendix}
\usepackage{lipsum}
\usepackage{MnSymbol}
\usepackage{algorithm}
\usepackage{multirow}
\usepackage[noend]{algpseudocode}
\usepackage{booktabs}
\usepackage{graphicx}
\usepackage{comment}

\newtheorem{theorem}{Theorem}
\newtheorem{proposition}[theorem]{Proposition}

\newtheorem{corollary}[theorem]{Corollary}
\theoremstyle{definition}
\newtheorem{definition}[theorem]{Definition}
\newtheorem{example}{Example}
\theoremstyle{remark}
\newtheorem{remark}{Remark}
\newcommand\norm[1]{\left\lVert#1\right\rVert}

\newcommand{\ide}{e}

\newcommand{\red}{\textrm{red}}

\newcommand{\intsym}{\hat{\phi}^H_t}
\newcommand{\intpoiss}{\hat{\phi}^{H^{\red}}_t}

\graphicspath{{figs/}}

\author[1]{Miguel Vaquero}
\author[2]{Jorge Cort\'es}
\author[3]{David Mart{\'\i}n de Diego}
\affil[1]{IE University (mvaquero@faculty.ie.edu)}
\affil[2]{University of California, San Diego (cortes@ucsd.edu)}
\affil[3]{ICMAT (david.martin@icmat.es)}

\title{Symmetry Preservation in Hamiltonian Systems: \\ Simulation and Learning}

\date{}
\begin{document}
\maketitle
\begin{abstract}

 This work presents a general geometric framework for simulating and learning the dynamics of Hamiltonian systems that are invariant under a Lie group of transformations. This means that a group of symmetries is known to act on the system respecting its dynamics and, as a consequence, Noether's Theorem, conserved quantities are observed. We propose to simulate and learn the mappings of interest through the construction of $G$-invariant Lagrangian submanifolds, which are pivotal objects in symplectic geometry. A notable property of our constructions is that the simulated/learned dynamics also preserves the same conserved quantities as the original system, resulting in a more faithful surrogate of the original dynamics than non-symmetry aware methods, and in a more accurate predictor of non-observed trajectories.  Furthermore, our setting is able to simulate/learn not only Hamiltonian flows, but any Lie group-equivariant symplectic transformation. Our designs leverage pivotal techniques and concepts in symplectic geometry and geometric mechanics: reduction theory, Noether's Theorem, Lagrangian submanifolds, momentum mappings, and coisotropic reduction among others. We also present methods to learn Poisson transformations while preserving the underlying geometry and how to endow non-geometric integrators with geometric properties. Thus, this work presents a novel attempt to harness the power of symplectic and Poisson geometry towards simulating and learning problems. 

\end{abstract}

\newpage
\tableofcontents

\section{Introduction}
 Hamiltonian systems are ubiquitous in physics and engineering, mainly because their mathematical properties make them powerful tools  for analyzing and designing complex systems. One key aspect is their ability to preserve energy, which is of pivotal importance in engineering as it ensures consistent behavior over time. Another paramount property of Hamiltonian systems is their ability to leverage the presence of symmetries, such as translational and rotational symmetry or
time invariance, to simplify their analysis and control. These symmetries offer insights
into the underlying system structure and facilitate the development of efficient control strategies. Due to the importance of symmetries in Hamiltonian systems, our ultimate goal here is to design algorithms able to simulate and learn Hamiltonian systems whilst respecting the underlying geometry and symmetries.

From a mathematical viewpoint, this paper aims to provide an algorithmic framework for generating all the $G$-equivariant symplectic transformations of a symplectic manifold. This framework equivalently provides a means to obtain transformations that conserve the corresponding momentum mappings. The solution to this task is achieved through the careful combination of various geometric constructions that rely on Lagrangian submanifolds, which are the main objects in symplectic geometry.  This believe is usually stressed by the so-called ``Symplectic Creed'':
\begin{quote}
  {\it ``Everything is a Lagrangian submanifold.''}
  
  \hfill {\it A.D. Weinstein}
\end{quote}
Due to their importance, Lagrangian submanifolds have been intensively studied and are quite well understood nowadays. Besides their theoretical importance, Lagrangian submanifolds are equipped with tools to describe them in easy terms, allowing their use in applications. Like the theory of generating functions or Morse families.

The results presented here are the key to advance two  fields: {\it geometric integrators} and {\it machine learning}. We describe next how our results fit into the aforementioned settings.

{\it $\bullet$ Geometric Integrators.} The field of geometric integration has become a well-established branch of numerical integration. Its goal is to simulate geometric dynamical systems, which are systems with underlying geometry, using surrogates that possess the same geometric properties as the original system. For example, a symplectic mapping is used to simulate a symplectic system. This methodology often leads to both qualitative and quantitative descriptions of the original model whose performance is superior to those provided by non-geometric schemes.

Motivated by the success of equipping an integrator with the same properties as the system it aims to describe, if the original system possesses symmetries (equivalently, conserved quantities as per Noether's Theorem), it becomes highly desirable to simulate the system using an integrator that also shares the same group of symmetries (or equivalently, the same conserved quantities). While there exist some geometric integrators known to conserve certain momentum mappings, we are currently unaware of a general procedure to generate all symplectic integrators that preserve the underlying symmetries. Therefore, the first question we address in this work is
\begin{center}
{\it How can we construct symplectic integrators that generate $G$-equivariant transformations?}
\end{center}
or equivalently
\begin{center}
{\it How can we construct symplectic integrators that conserve the momentum mapping?}
\end{center}

{\it $\bullet$  Machine Learning.} 
The modeling of dynamical processes in various applications necessitates a thorough understanding of the phenomena under analysis. While models are highly accurate in many cases, they face significant challenges when applied to complex dynamical systems, such as climate dynamics, brain dynamics, biological systems, or financial markets. Fortunately, by incorporating machine learning techniques, we can harness the power of data-driven learning algorithms to enhance prediction and control of these systems. Consequently, we have witnessed a growing body of research that combines the fields of dynamical systems and machine learning in recent years. This interdisciplinary approach serves two purposes: firstly, using dynamical systems to gain deeper insights into machine learning, such as designing or analyzing optimization algorithms or integrating structures from dynamical systems to facilitate the learning process. Secondly, employing machine learning to aid in the understanding of complex dynamical systems by leveraging data. 

Another recent trend in the machine learning/dynamical systems community focuses on combining the successful aspects of neural networks with other known structures to enhance their performance. For example, when dealing with systems that possess underlying structures such as Lagrangian or Hamiltonian properties, one can directly identify the equations of motion or determine the Hamiltonian/Lagrangian that governs them. This straightforward concept has given rise to the development of Hamiltonian Neural Networks (HNN) and Lagrangian Neural Networks (LNN). These advancements have led to improved prediction accuracy and better models, garnering significant attention from practitioners in the machine learning field.

As previously stated, in classical mechanics the study of symmetries and momentum mappings has become one of the recurrent topics due to its importance in applications. Therefore, when learning geometric systems, a question of pivotal importance is: 
\begin{center}
{\it How can we design learning paradigms for Hamiltonian systems that preserve the underlying symmetries and use them to improve the performance of the resulting models?  
}    
\end{center}

The aforementioned questions are significant in their own right and certainly merit theoretical investigation. However, it is important to highlight that they have broader downstream implications, for which the present paper serves as a starting point. An intriguing future research direction would involve addressing the challenge of designing policies for controlling systems in various frameworks, such as robotics with rotational components. This direction is also motivated by the successful utilization of geometric tools in solving optimal control problems.

\subsection{Literature Review}

{\it Geometric Integration.} The area of geometric integration is supported nowadays by a vast community of researchers and it would be impossible to summarize all relevant contributions. Classic references are the books~\cite{hairer,McLachlan6}, which cover topics ranging from symplectic to Poisson integration and their properties. Generating functions have been extensively used in symplectic integrators~\cite{leok}. Some recent works focused on the simulation of learned Hamiltonians~\cite{sina2}. 

Regarding Poisson integrators, a recent survey can be found in~\cite{david}. Some references~\cite{Ge,Ge-Marsden} are particularly important for this work and are based on the use of generating functions to describe Hamiltonian dynamics~\cite{cosserat,FLMV,JAY}.

{\it Machine Learning and Hamiltonian Dynamical Systems.} The machine learning literature is very recent, but already vast and increasing at a high pace. Among the initial works we would like to mention the results in \cite{hamiltonian_neural_networks}, where Hamiltonian Neural Networks (HNN) where introduced. In \cite{lagrangian_neural_networks} the Lagrangian counterpart of HNN was developed and more recently even a dissipative version appeared \cite{dissipative_neural_networks}. Several refinements and improvements showed up shortly after the introduction of HNN, like the inclusion in the learning process of recurrent neural networks in \cite{symplectic_recurrent_neural_networks}, and also the use of other architectures like the ones in \cite{symplectic_karniadakis}.  The use of Lagrangian submanifolds to learn Hamiltonian dynamical systems has  been treated in \cite{chen-tao}, where the authors use Type II generating functions to parameterize the Lagrangian submanifolds of interest.

    The inclusion of symmetry in the learning process of Hamiltonian systems has not been extensively treated yet, to the best of the authors' knowledge. Some related works include~\cite{reszende}, where Poisson brackets are used in an attempt to obtain symmetric mappings. The results in~\cite{ifac_symmetries} use a penalty for the violation of momentum conservation. The recent works~\cite{Sina,eldred2023liepoisson} also treats the problem of identifying symmetries for better learning procedures. Regarding Poisson geometry, the only work we are aware of is the results in the recent paper~\cite{eldred2023liepoisson} and~\cite{poisson_karniadakis}. 
    
    Finally, we would like to mention that the line of research presented here fits into the recent attempts to endow learning procedures with physical information~\cite{pinns_karniadakis}, where modelling and data are combined to improve the learning and prediction processes.

\subsection{Statement of Contributions}

%
%
%
%

{\it Theoretical Contribution.} We introduce a comprehensive framework that enables the generation of all $G$-equivariant symplectic transformations for specific symplectic manifolds, leveraging the properties of Lagrangian submanifolds. Equivalently, our framework allows for the generation of all symplectic transformations that preserve momentum. Our presentation emphasizes the significance of symplectic groupoids in comprehending and characterizing Hamiltonian systems, opening avenues for further generalizations to complex structures such as singular actions or non-linear Poisson structures.

{\it Applications.} 
Building upon our theoretical constructions, we introduce a reduction and reconstruction theory for $G$-invariant Lagrangian submanifolds. This theory enables the development of equivariant integrators and a learning paradigm that respects symmetries. Additionally, we outline the methodology for learning Poisson transformations by utilizing the theory of generating functions while preserving the underlying geometry. All our findings are carefully illustrated using the rigid body as a benchmark. We compare the proposed simulation methods with and our learning methods with state of the art methodologies (see~\cite{chen-tao}).

 Lastly, we present several ongoing research directions. Like a framework that facilitates the integration of geometric properties into non-geometric integrators, effectively ``geometrizing" them. Or the design of new Poisson integrators by solving the Hamilton-Jacobi equation using machine learning techniques.

\section{Notation and Background}

In this paper all manifolds and mappings are infinitely
differentiable ($C^{\infty}$). The cotangent bundle of a manifold $M$ is denoted by $T^*M$ and the canonical projection is denoted by $\pi_M : T^*M\rightarrow M$. Given a map $f:M\rightarrow N$ between
manifolds $M$ and $N$, we use the notation $Tf$ to denote the
tangent map  ($Tf:TM\rightarrow
TN)$, and $Tf(q)$, where $q$ is a point on $M$, to denote the tangent
map at that point $Tf(q):T_qM\rightarrow T_{f(q)}N$. We use $graph(f)$ to denote the set $graph(f) = \{(q,f(q))\in M\times N \textrm{ such that } q \in M\}$. The evaluation of a vector field at a point $q\in M$
 reads $X(q)$, we  use analogous notation for differentiable forms. The flow  of the vector fields under consideration is assumed to be defined globally, although our results hold for locally defined flows with the obvious modifications. Throughout this paper $G$ is a connected Lie group and $\mathfrak{g}$ the corresponding Lie algebra. Given an action of the Lie group $G$ on the manifold $M$, $\Phi:G\times M\rightarrow M$, and $q\in M$ then $Orb(q)$ represents the set $\{\Phi(g,q)\textrm{ such that }g\in G\}\subseteq M$.

\paragraph{$\bullet$ Symplectic, Lagrangian, and Poisson Manifolds:}  We introduce here the basic geometric structures used along the paper. See also~\cite{AM87,marsden3} for a complete description of these topics.
\begin{definition}[Symplectic Manifold]
    A symplectic manifold is a tuple $(M,\omega)$, where $M$ is a differentiable manifold and $\omega$ is a non-degenerate closed two-form. Given a symplectic manifold and a function over it, $H: M \rightarrow \mathbb{R}$, the corresponding Hamiltonian  vector field, $X_H$, is defined by $i_{X_H}\omega = dH$. A symplectic transformation between two symplectic manifolds, $(M,\omega_M)$ and $(N,\omega_N)$, is a mapping $f: M \rightarrow N$ such that $f^*\omega_N = \omega_M$. Symplectic diffeomorphism are called symplectomorphisms. 
\end{definition}

\begin{example}[Canonical Symplectic Form in the Cotangent Bundle]
    The cotangent bundle of a manifold, say $T^*M$, is endowed with a canonical symplectic form by taking $\omega_M = -d\theta$, where $\theta$ is the Liouville one form. In local coordinates $(x^i,p_i)$ this form takes the familiar expression $dx^i \wedge dp_i$. The opposite of the canonical symplectic structure on the cotangent bundle, that is $(T^*M,-\omega_M)$ is denoted by $T^*M^-$.
\end{example}

\begin{definition}[Lagrangian Submanifold]
    A Lagrangian submanifold ${\mathcal L}$ of a symplectic manifold $(M,\omega)$ is an embedded submanifold such that the restriction of $\omega$ to ${\mathcal L}$ is zero, that is, for all $q\in M$ and $u_q, v_q\in T_p{\mathcal L}$, $\omega(q)(u_q, v_q)=0$. By algebraic considerations, symplectic manifolds have even dimension and Lagrangian submanifolds have half of the dimension of the symplectic manifold they live in.
\end{definition}

\begin{example}[Type I Generating Functions] Given a manifold $M$ and its cotangent bundle $T^*M$, any differentiable function $S:M \rightarrow \mathbb{R}$ produces a Lagrangian submanifold in $T^*M$ by just taking $graph(dS)\subset T^*M$. In canonical coordinates, $(x^i,p_i)$, this submanifold is just given by $(x^i, p_i=\displaystyle\frac{\partial S}{\partial x^i}(x))$, where $x\in M$.
  \end{example}

\begin{definition}[Poisson Manifold]
    A {\bf Poisson structure} on a  differentiable manifold $P$ is given by a bilinear map
\[
\begin{array}{rcc}
C^{\infty}(P)\times C^{\infty}(P)&\longrightarrow& C^{\infty}(P)\\
(f, g)&\longmapsto& \{f, g\}
\end{array}
\]
called the {\bf Poisson bracket}, satisfying the following properties: 
\begin{itemize}
\item[(i)] \emph{Skew-symmetry},  $\{g, f\}=-\{f, g\}$;
\item[(ii)] \emph{Leibniz rule}, $\{fg, h\}=f\{g, h\}+g\{f, h\}$; 
\item[(iii)] \emph{Jacobi identity},  $\{\{f, g\}, h\}+\{\{h, f\}, g\}+\{\{g, h\}, f\}=0$;
\end{itemize}
for all $f, g, h\in C^{\infty}(P)$. 
\end{definition}
If $P$ is a manifold and $\{\,,\,\}$ a Poisson structure on $P$, then the pair $(P, \{\,,\,\})$ is a Poisson manifold.
\begin{example}[Dual of a Lie Algebra, $\mathfrak{g}^*$]
    If ${\mathfrak g}$ is a Lie algebra with Lie bracket $[\; ,\; ]$, then it is defined a   Poisson bracket on
 ${\mathfrak g}^*$  by 
 \[
 \{\xi, \eta \}(\alpha) = -\langle\alpha ,[\xi,\eta]\rangle\; ,
 \] where $\xi$ and $\eta \in
{\mathfrak g}$ are equivalently considered as linear forms on ${\mathfrak g}^*$, and $\alpha \in {\mathfrak g}^*$.
This  linear Poisson structure on ${\mathfrak g}^*$ is called the
{\it Kirillov-Kostant-Souriau Poisson structure}. 
\end{example}
\paragraph{$\bullet$ Actions, Lifted Actions, and Co-adjoint Actions:} See~\cite{marsden3} for a complete description of these concepts. Let $G$ be a connected Lie group acting freely and properly on a manifold $M$ by a left action $\Phi$,
\[
\begin{array}{rccl}
\Phi:&G\times M&\longrightarrow &M \\\noalign{\medskip}
& (g,q) &\longmapsto & \Phi(g,m)=g\cdot q
\end{array}
\]
Given $g\in G$, we denote by $\Phi_g:M \rightarrow M$
the diffeomorphism defined by $\Phi_g(q)=\Phi(g,q)=g\cdot q$. Recall that if the action $\Phi$ is free and proper (see \cite{marsden3}, Proposition 4.1.23) the quotient $M/G$ can be endowed with a manifold structure such that the canonical projection $\pi:M\rightarrow M/G$ is a $G$-principal bundle. The action $\Phi$ introduced above can be lifted to actions on the tangent and cotangent bundles, $\Phi^T$ and $\Phi^{T^*}$  respectively. We briefly recall here their definitions.

{\it - Lifted action on $TM$}. We introduce the action $\Phi^T:G\times TM\rightarrow TM$
such that $\Phi_{g}^{T}:TM\rightarrow TM$ is defined by
\[
\Phi_g^{T}(v_q)=T\Phi_g(q)(v_q)\in T_{gq}M
\quad 
\textrm{ for } v_q\in T_qM.
\]

{\it - Lifted action on $T^*M$}. Analogously, we introduce the following  action $\Phi^{T^*}:G\times T^*M\rightarrow T^*M$ such that $\Phi_{g}^{T^*}:T^*M\rightarrow T^*M$ is defined by
\[
\Phi_g^{T^*}(\alpha_q)=(T\Phi_{g^{-1}})^*(gq)(\alpha_q)\in T_{gq}^*M \quad\textrm{ for } \alpha_q\in T_q^*M,
\]
that is,
\[
\langle \Phi_g^{T^*}(\alpha_q), v_{gq}\rangle=\langle \alpha_q, (T\Phi_{g^{-1}})(gq)(v_{gq})\rangle \quad\textrm{ where } v_{gq}\in T_{gq}M.
\]
 Both actions can be easily checked to be free and proper. If $\alpha_q\in T^*M$, we denote the orbit through $\alpha_q$ by
$Orb(\alpha_q)$. We  make use  of $Ad^*$ to represent the Coadjoint action on the dual of $\mathfrak{g}$ given by
\[
\begin{array}{rccl}
Ad^*:&G\times \mathfrak{g}^*&\longrightarrow& \mathfrak{g}^* \\ \noalign{\medskip}
& (g,\mu)&\longmapsto & Ad^*_{g}(\mu)=\mu\circ TR_{g}\circ TL_{g^{-1}},
\end{array}
\]
where $L_g(h)=g\cdot h$ and $R_g(h)=h\cdot g$ are the left and right
multiplication on the group $G$. Notice that the Coadjoint action is a
left action. Given $\mu\in\mathfrak{g}^*$, $Orb^{Ad^*}(\mu)$ denotes the orbit by the Coadjoint action through $\mu$.

\begin{definition}[$G$-Equivariant and $G$-Invariant Transformations] Let $G$ be a Lie group acting on $M$ and $N$. Then, a mapping $f: M \rightarrow N$ is said equivariant if $f(g\cdot{}q) = g\cdot{}f(q)$ for all $q\in M$. A mapping $f:M \rightarrow \mathbb{R}$ is said $G$-invariant when $f(g\cdot{}q) = f(q)$, that is, the function $f$ is constant on the orbits of the action of $G$ over $M$. When the Lie group acting  $G$ is clear, we  use terms equivariant or invariant map without reference to $G$.
\end{definition}
\paragraph{$\bullet$ Momentum Mappings:} It is well-known, see \cite{marsden3}, that there exists a $G$-e\-qui\-var\-iant momentum mapping for the
lifted action on $T^*M$ with respect to its \textrm{canonical} symplectic form, from now on denoted by
$\omega_M$. This momentum map is given by $J:T^*M\rightarrow\mathfrak{g}^*$, 
where $J(\alpha_q)$ is such that
$J(\alpha_q)(\xi)=\alpha_q(\xi_{M}(q))$ for $\xi\in\mathfrak{g}$. Here $\xi_M$ is the vector field  on $M$ determined via the action $\Phi$, called the infinitesimal generator. The integral curve of $\xi_M$ passing through
  $q\in M$ is  $t\rightarrow exp (t\xi)\cdot{}q$, where $exp: \mathfrak{g} \rightarrow G $ is the exponential mapping of a Lie group. See~\cite{marsden3} for a complete description of these concepts.

Given $\xi\in\mathfrak{g}$, we denote by $J_{\xi}:T^*M\rightarrow
\mathbb{R}$ the real function obtained by the natural pairing between
$\mathfrak{g}$ and $\mathfrak{g}^*$, $J_{\xi}(\alpha_q)=\langle
J(\alpha_q),\xi\rangle$. By the definition of momentum mapping, we have 
$\xi_{T^*M}=X_{J_{\xi}}$,
where $\xi_{T^*M}$ is the fundamental vector field generated by $\xi$
via the action $\Phi^{T^*}$. Indeed, we have
                             $i_{\xi_{T^*M}}\omega_M=dJ_{\xi}$ and $X_{J_\xi}$ is the Hamiltonian vector field
  for the function $J_\xi$, that is, $i_{X_{J_\xi}}\omega_M=d J_\xi$.

  In the case where the configuration manifold is a Lie group, $G$, then if we are considering the (lifted) left action of $G$ on $T^*G$, $J_L$ is the corresponding momentum mapping. Equivalently, if we consider the right action of $G$ on $T^*G$ then $J_R$ is a momentum mapping. Formally,
\[
\begin{array}{rccl}
J_L&:T^*G&\longrightarrow &\mathfrak{g}^*\\\noalign{\medskip}
&\alpha_{g}&\longmapsto & \langle J_L(\alpha_g),\xi\rangle=\langle\alpha_g,T_\ide R_g(\xi)\rangle,
\end{array}
\]
 and 
\[
\begin{array}{rccl}
J_R:&:T^*G&\longrightarrow &\mathfrak{g}^*\\\noalign{\medskip}
&\alpha_{g}&\longmapsto &\langle J_R(\alpha_g),\xi\rangle=\langle \alpha_g,T_\ide L_g(\xi)\rangle.
\end{array}
\]
In the left trivialization of $T^*G\equiv G\times {\mathfrak g}^*$, we have that 
  $$
    J_L(g, \mu)=Ad^*_g\mu\, , \qquad  
    J_R(g, \mu)=\mu 
.$$  

    

\paragraph{$\bullet$ Coisotropic Reduction}\label{coisotropic_reduction} The next result, combined with the fact that $\Phi^{T^*}$ is free
and $G$ connected, ensures that, for a connected Lie group, every
$\mu\in \mathfrak{g}^*$ is a regular value and so $J^{-1}(\mu)$ is a
submanifold. In fact, the next proposition characterizes regular values of
momentum mappings taking into account the infinitesimal behavior of
the symmetries. We define $\mathfrak{g}_q=\{\xi\in\mathfrak{g}$
     such that $\xi_M(q)=0\}$ and $q\in M$.

 \begin{proposition}[See \cite{ham_stages}, Prop. $1.1.2$] Let $(M,\
  \Omega)$ be a symplectic manifold and $G$ a Lie group which acts by
  symplectomorphism with equivariant momentum map $J$. An element
  $\mu\in\mathfrak{g}^*$ is a regular value of $J$ iff 
  $\mathfrak{g}_p =\{0\}$ for all $p\in J^{-1}(\mu)$.
\end{proposition}

\begin{remark} {\rm In the case that concerns us, namely $(T^*M, \ \omega_M)$
  with the action $\Phi^{T^*}$, this result implies that $J^{-1}(\mu)$ is always a submanifold of $T^*M$.}
\end{remark}

\begin{theorem}[Coisotropic Reduction. See \cite{Weinstein}, p. $12$]\label{reductionlagrangian}
Let $(M,\omega)$ be a symplectic manifold,  $C\subset  M$ a
coisotropic submanifold and $C/\hspace{-1.5mm}\sim$ the quotient space of $C$ by the
characteristic distribution $D=\textrm{ker}(\omega_{|C})$; we shall
denote by $\pi:C\rightarrow C/\hspace{-1.5mm}\sim$ the canonical projection and by
$\omega_C$ the natural projection of $\omega$ to $C/\hspace{-1.5mm}\sim$ (notice that
$(C/\hspace{-1.5mm}\sim,\ \omega_C)$ is again a symplectic manifold,
assuming that it is again a manifold). Assume that $L\subset M$ is a Lagrangian submanifold such that $L\cap
C$ has clean intersection, then $\pi(L\cap C)$ is a Lagrangian
submanifold of $(C/\hspace{-1.5mm}\sim,\ \omega_C)$. The following diagram illustrates the above situation
\[
\xymatrix{
L\cap C \;\ar@{^{(}->}[rr]^{i_{L\cap C}}\ar[dd]^{\pi}&& C \;\ar@{^{(}->}[rr]^{i_{C}}\ar[dd]^{\pi}&& M \\ \\
\pi(L\cap C)\; \ar@{^{(}->}[rr]_{i_{\pi(L\cap C)}}&& C/\sim & 
}
\]

\end{theorem}

The last two results combined permit the reduction of Lagrangian submanifolds using $J^{-1}(\mu)$ as $C$ in Theorem~\ref{reductionlagrangian}.

\section{Hamiltonian Systems with Symmetry:  Challenges in Simulation and Learning}\label{challenges}
The following result characterizes the key properties of Hamiltonian systems with symmetry, which we aim to replicate when simulating or learning such systems. To simplify the presentation, we introduce our results in the case where the configuration manifold is a Lie group. However, it is important to note that our results are applicable in the more general setting of a lifted action. We also assume that $G$ acts on $T^*G$ on the left, and therefore $J_L$ is the corresponding momentum mapping.

\begin{proposition}[Hamiltonian Systems with Symmetry]\label{prop_symmetry}
Given a Hamiltonian $H:T^*G \rightarrow \mathbb{R}$ which is $G$-invariant, then the following properties hold:

\begin{enumerate}
    \item The Hamiltonian flow, denoted by $\phi^H_{t}$, is $G$-equivariant;
    \item $J_L$ is conserved throughout the trajectories of the systems, that is,\\ $J_L(\phi^H_t(\alpha_g)) = J_L(\alpha_g)$;
    \item There is a reduced Hamiltonian $H^{\red}: \mathfrak{g}^* \rightarrow \mathbb{R}$ such that $H^\red \circ J_R = H$. The Hamiltonian $H^\red$ generates a Poisson dynamics on $\mathfrak{g}^*$ The flows of the corresponding Hamiltonian vector fields make the diagram in Figure~\ref{fig:1} commutative, where $\pi$ is the projection over the quotient $T^*G/G \equiv \mathfrak{g}^*$, which amounts to $J_R$.
\begin{figure}[H]
    \centering
\[
\xymatrix{
T^*G\ar[rr]^{\phi^H_t} \ar[dd]^{\pi = J_R}&& T^*G \ar[dd]^{\pi = J_R}\\ \\
\mathfrak{g}^*\ar[rr]^{\phi^{H^\red}_t} && \mathfrak{g}^*
}
\]
    \caption{Illustration of how the equivariant and the reduced transformation match, making the diagram commutative.}
    \label{fig:1}
\end{figure}
\end{enumerate}
\end{proposition}

 The properties mentioned above play a crucial role in determining the qualitative behavior of the system and setting it apart from other dynamics. Therefore, in order to accurately approximate or learn the mapping $\phi^H_{\Delta t}$  , it becomes imperative to search for transformations from $T^*G$ to $T^*G$  that also possess these properties of the original dynamics. In the following, we outline the primary challenges associated with achieving these properties when designing geometric integrators or when learning the dynamics.

\subsection{Simulation of Systems with Symmetries}\label{sec:simulation_of_systems_with_symmetries}
The first problem we address is the construction of numerical methods capable of simulating systems with symmetry while respecting the properties outlined in  Proposition~\ref{prop_symmetry}.

\paragraph{Challenges:} We seek to design pairs of integrators\footnote{By integrator we mean a mapping that approximates the flow of a vector field at a fixed time step. When the mapping is symplectic, we say the integrator is symplectic. When the mapping is Poisson, we say the integrator is Poisson.} approximating $\phi^H_t$ ( denoted $\intsym$) and $\phi^{H^\red}_t$ (denoted $\intpoiss$) satisfying:
\begin{itemize}
    \item $\intsym$ is symplectic and $\intpoiss$ is Poisson, meaning they are respectively symplectic and Poisson diffeomorphisms.
    \item The integrator approximating $\phi^H_t$, $\intsym$ conserves the momentum mapping, $J_L$, and is equivariant.
    \item $\intsym$ and $\intpoiss$ approximate both $\phi^H_t$ and $\phi^{H^\red}_t$ making the diagram analogous to the one in Figure~\ref{fig:1} commutative. That is $ J_R\circ\intsym = \intpoiss \circ J_R$.
\end{itemize}

\subsection{Learning Systems with Symmetries}\label{sec:learning_systems_with_symmetry}
The second problem of interest for this work is somehow complementary to the simulation problem. In the learning problem, we are given a set of trajectories and we would like to learn the mechanism that generates them. More precisely, given a data-set of pairs $D = \{(\alpha'_{g'},\alpha''_{g''})_i,\ i \in I\}$ of a Hamiltonian system evolving on a Lie group, where the momentum mapping is preserved ($J_L(\alpha'_{g'}) = J_L(\alpha''_{g''})$), we would like to learn the $G$-equivariant transformation $\phi^H_{\Delta t}$ that generates the data ($\phi^H_{\Delta t}(\alpha'_{g'}) = \alpha''_{g''}$).

\paragraph{Challenges:} We would like to use a neural network to obtain, through training on the data set $D$, an approximation of the mapping $\phi^H_{\Delta t}$ (denoted $\intsym$) satisfying:

\begin{subequations}
\begin{flalign}
  \tag{{\it$L$-symp}}
 & \text{$\bullet$ $\intsym$ should be symplectic, that is, conserve the symplectic form.}
  \label{item:1}
  \\ 
  \tag{{\it $L$-equivariant}}
& \text{$\bullet$ $\intsym$ should conserve the momentum mapping and be equivariant.}
  \label{item:2}
  \end{flalign}
\end{subequations}


\section{Equivariant Transformations Through Invariant Submanifolds}


In this section, we present the key observations that contribute to solving the main challenges outlined in Section~\ref{challenges}. Our approach is based on the significance of Lagrangian submanifolds in symplectic geometry \cite{ AM87,GuiStern,LibMa,Weinstein}. Our strategy hinges upon two main ideas:
  \begin{itemize}
  \item Transform the problems of simulating and learning the dynamics into a problem of finding Lagrangian submanifolds.
    \item Employ the theory of generating functions to effectively track the Lagrangian submanifolds of interest and identify the one that best suits the problem at hand.
  \end{itemize}
As stated, equivariant symplectic transformations correspond to invariant Lagrangian submanifolds.
 
\begin{definition} Let  $(T^*G,\omega_G)$ be a symplectic manifold endowed
   with a Hamiltonian lifted action $\Phi$. A Lagrangian submanifold ${\mathcal L} \subset T^*G$ is a {\it $G$-invariant
  Lagrangian submanifold} if it is invariant by all the elements of the action, i.e.,  we have $\Phi^{T^*}_g({\mathcal L})={\mathcal L}$ for all $g\in G$.
\end{definition}

The main motivation to study these objects is the following result, which is the equivariant analogue of the fact that symplectic mappings correspond to Lagrangian submanifolds in $T^*G^- \times T^*G$. Roughly speaking, equivariant symplectic mappings correspond to Lagrangian submanifolds in $T^*G$.

\begin{proposition}[$G$-equivariant Mappings as Lagrangian Submanifolds] Let $f: T^*G \rightarrow T^*G$, where $G$ acts through a lifted action on $T^*G$. Then, $f$ is $G$-equivariant symplectomorphism if and only if $graph(f)\subset T^*G^-\times T^*G$ is a $G$-invariant Lagrangian submanifold.
\end{proposition}

\proof
It is a well-known fact (\cite{marsden3}) that a symplectic mapping $f:T^*G \rightarrow T^*G$ generates a Lagrangian submanifold by means of $graph(f)\subset T^*G^- \times T^*G$. So we only need to care about proving equivariance.  Assume $f$ is equivariant. Then, for any $h\in G$, $\alpha_g\in T^*G$, from the computation
\begin{align*}
h\cdot{} (\alpha_{g},f(\alpha_g)) = (h\cdot{}\alpha_g,h\cdot{}f(\alpha_g)) = (h\cdot{}\alpha_g,f(h\cdot{}\alpha_g))    
\end{align*}
we deduce that $graph(f)$ is a $G$-invariant Lagrangian submanifold. Reversing the computations yields the equivalence.

\qed

The preceding result primarily converts the task of searching for requirements \eqref{item:1} and \eqref{item:2} into the quest for invariant Lagrangian submanifolds. This observation, far from being purely theoretical, enables us to leverage the powerful tools associated with Lagrangian submanifolds to address the problems presented in this paper. Fortunately, $G$-invariant Lagrangian submanifolds possess robust geometric properties that facilitate their description. The next result, which characterizes $G$-invariant Lagrangian submanifolds, subsequently leads to the implications of Noether's Theorem.

 \begin{proposition}\label{lemma} Under the previous assumptions, we have:
 \begin{enumerate}
     \item  Let ${\mathcal L}\subset T^*G$ be a  Lagrangian submanifold of
 $(T^*G, \ \omega_G)$. Then $J_L$ is constant along ${\mathcal L}$ if and only if ${\mathcal L}$ is $G$-invariant.
 \item If a G-invariant Lagrangian submanifold lives in a level set of the momentum mapping, ${\mathcal L}\subset J_L^{-1}(\mu)$, then the isotropy group is $G$, that is $G_\mu = G$.
 \item If $\mu$ is such that $G_{\mu}=G$, then $J_L^{-1}(\mu)$ is a
  coisotropic submanifold of $T^*G$.
 \end{enumerate}
\end{proposition}

 \proof $ $\newline
 \begin{enumerate}
 \item Assume first  that ${\mathcal L}$ is $G$-invariant.  Let  $\alpha_q\in {\mathcal L}$ and $X\in T_{\alpha_g}{\mathcal L}$, then 
\begin{equation}\label{one}
 d(J_L)_{\xi}(\alpha_g)(X)=(i_{\xi_{T^*G}}\omega_G)(\alpha_g)(X)=\omega_G(\alpha_q)(\xi_{T^*G}(\alpha_g),X).
\end{equation}
Now, notice that
\[
\xi_{T^*G}(\alpha_q)=\textrm{tangent vector at $t=0$ to the curve $exp(t\xi)(\alpha_q)$.}
\]

Since $exp(t\xi)(\alpha_q)$ is contained in the orbit of $\alpha_q\in {\mathcal L}$,
and $Orb(\alpha_q)\subset {\mathcal L}$ since ${\mathcal L}$ is $G$-invariant (that is, $G\cdot
{\mathcal L}\subseteq {\mathcal L}$), we deduce that $\xi_{T^*G}(\alpha_q)\in T_{\alpha_q}{\mathcal L}$.
Therefore, \eqref{one} vanishes since ${\mathcal L}$ is Lagrangian. Finally, since
$(J_L)_{\xi}$ is constant along ${\mathcal L}$, we have $(J_L)_{\xi}(\alpha_g)=c_\xi$ for all
$\alpha_g\in {\mathcal L}$ and for all $\xi\in\mathfrak{g}$ and thus,
$J_L(\alpha_g)=\mu$ for all $\alpha_g\in L$ (such that $\mu(\xi)=c_\xi$). Reversing the computations we obtain the other implication.

\item 
Since the momentum mapping $J_L$ is equivariant with respect to the
  $Ad^*$ action, then the result follows.

\item This follows from \cite{marsden3}.
  \end{enumerate}

\qed

\vspace*{0.5cm}

Next, we introduce some auxiliary constructions that will be useful to state  the main results of this work. Given the natural (left) action of a Lie group on itself $\Phi(g,g') = g\cdot{}g'$, we can associate the diagonal action
\[
\begin{array}{rccl}
\Phi_e:&G\times \left(G\times G\right)&\longrightarrow &G\times G\\\noalign{\medskip}
&(g,(g',g''))&\rightarrow& (g\cdot{}g',g\cdot{}g'').
\end{array}
\]
The associated cotangent lifted action is easily seen to be
$\Phi_e^{T^*}(g,(\alpha'_{g'},\alpha''_{g''}))=(\Phi^{T^*}_g(\alpha'_{g'}),\Phi^{T^*}_g(\alpha''_{g''}))$. The corresponding
momentum mapping is given by
\[
\begin{array}{rccl}
J_e:&T^*(G\times G)&\longrightarrow &\mathfrak{g}^*\\\noalign{\medskip}
&(\alpha'_{g'},\alpha''_{g''})&\rightarrow& J_L(\alpha'_{g'}) +J_L(\alpha''_{g''}).
\end{array}
\]
The {\it twist} map  bridges the manifold $T^*G^-\times T^*G$ with the cotangent bundle $T^*(G\times G)$
\[
\begin{array}{rccl}
twist:& T^*G^-\times T^*G& \longrightarrow& T^*G\times T^*G \equiv T^*(G\times G)\\ \noalign{\medskip}
& (\alpha'_{g'},\alpha''_{g''})&\rightarrow& (-\alpha'_{g'},\alpha''_{g''}).
\end{array}
\]
Notice that this mapping is a symplectomorphism and therefore sends Lagrangian submanifolds in $T^*G^-\times T^*G$ to Lagrangian submanifolds in $T^*(G\times G)$.

Finally,  the next two mappings are useful to construct Poisson transformations through Lagrangian submanifolds in $T^*G$,
\[
\begin{array}{rccl}
p_1^-:& T^*(G\times G)& \longrightarrow& T^*G\\ \noalign{\medskip}
& (\alpha'_{g'},\alpha''_{g''})&\longmapsto& -\alpha'_{g'},
\end{array}
\]
\[
\begin{array}{rccl}
p_2:& T^*(G\times G)& \longrightarrow& T^*G\\ \noalign{\medskip}
& (\alpha'_{g'},\alpha''_{g''})&\longmapsto& \alpha''_{g''}.
\end{array}
\]

\begin{corollary}\label{equivariant}
\begin{enumerate} \item Let ${\mathcal L}$ be a  Lagrangian submanifold of
 $T^*G^-\times T^*G$. Then $J_e$ is constant along ${\mathcal L}$ if and only if
 ${\mathcal L}$ is $G$-invariant. That is ${\mathcal L}$ is $G$-invariant if and only if
 ${\mathcal L}\subset  J^{-1}_e(\mu)$ for some $\mu\in\mathfrak{g}^*$.

\item Since the momentum mapping $J_e$ is equivariant with respect to the
  $Ad^*$ action, then the isotropy group of $\mu$ in  item $1$   with respect to this
  action is the whole group, $G_\mu=G$. 

\item If $\mu$ is such that $G_{\mu}=G$, then $J^{-1}_e(\mu)$ is a
  coisotropic submanifold.
\end{enumerate}
\end{corollary}

The proof of the three statements is a straightforward consequence of Proposition~\ref{lemma}. The next result is the main theoretical contribution of the paper. It bridges equivariant transformations with Lagrangian submanifolds in a reduced space. This observation, jointly with techniques like generating functions to describe locally Lagrangian submanifolds, allows to parameterize the transformations of interest.

\begin{theorem}[Equivariant Mappings as Invariant Submanifolds]\label{equivariant_lagrangian} There is an explicit one-to-one correspondence between equivariant, momentum preserving, symplectic transformations $f: T^*G\rightarrow T^*G$ and Lagrangian submanifolds  ${\mathcal L}^{red} \subset T^*G$. This correspondence can be explicitly described by
\begin{center}
     $f(\alpha'_{g'}) = \alpha''_{g''}$ if and only if $(g')^{-1}\alpha''_{g''} \in {\mathcal L}^{red}$.
\end{center}
\end{theorem}

\proof For the sake of clarity of the exposition, we divide the proof into two steps.
\begin{itemize}
\item {\it First step:}
The first step is to notice that there is a correspondence between equivariant symplectic transformations conserving the momentum mapping, say $f:T^*G\rightarrow T^*G$, and Lagrangian submanifolds ${\mathcal L}$ in $T^*(G\times G)$ living in $J_e^{-1}(0)$. Given $f$, the mentioned correspondence associates to it the image of $graph(f)$ through the {\it twist} mapping previously introduced. More precisely,
\[
graph(f) \subset T^*G^-\times T^*G \Leftrightarrow {\mathcal L} = twist(graph(f)) \subset T^*(G\times G).
\]
We can conclude that for $(-\alpha'_{g'},\alpha''_{g''})\in {\mathcal L}$ we have $J_e(-\alpha'_{g'},\alpha''_{g''})=0$, because
\[
J_e(-\alpha'_{g'},\alpha''_{g''}) = 0 \Leftrightarrow J_L(-\alpha'_{g'})+J_L(\alpha''_{g''})=0\Leftrightarrow J_L(\alpha'_{g'})=J_L(\alpha''_{g''}).
\]
Therefore, conservation of the momentum mapping by $f$ is equivalent to ${\mathcal L}\subset J_e^{-1}(0)$. Observe that the mapping {\it twist} is a symplectomorphism, and allows us to transform the graph of $f$ into a Lagrangian submanifold in the cotangent bundle $T^*(G\times G)$ where cotangent bundle reduction applies.

\item {\it Second step:}
Here, we prove that there is a correspondence between Lagrangian submanifolds inside $J_e^{-1}(0)$ and Lagrangian submanifolds in $T^*G$. This is due by coisotropic reduction, since by our previous results $J_e^{-1}(0)$ is a coisotropic manifold and ${\mathcal L}\subset J_e^{-1}(0)$ we obtain that there is a Lagrangian submanifold in $\displaystyle\frac{J_e^{-1}(0)}{G}$ that we denote by ${\mathcal L}^{\red}$. The identification 
\[
\displaystyle\frac{J_e^{-1}(0)}{G} \equiv T^*\left(\frac{G\times G}{G}\right) \equiv T^*G
\]
is due to cotangent bundle reduction, see \cite{ham_stages}. In what follows we just make explicit this correspondence. Assume that $(-\alpha^1_{g^1},\alpha^2_{g^2})\in {\mathcal L}$ and ${\mathcal L}$ is $G$--equivariant.  Now, remember that by the definition of cotangent bundle reduction by the following rule
\[
[(-\alpha'_{g'},\alpha''_{g''})]\in \displaystyle\frac{J_e^{-1}(0)}{G}\leftrightarrow \beta_{(g')^{-1}g''}\in T^*\left(\displaystyle\frac{G\times G}{G}\right) \equiv T^*G
\]
where if $X_{g''}\in  T_{g''}G$ then 
\[
\langle \beta_{(g')^{-1}g''}, T_{g''}L_{g'}^{-1}(X_{g''})\rangle =
\langle \alpha_{g''}, X_{g''} \rangle.
\]
The last expression can be rewritten as
\[
\alpha''_{g''}=g'\cdot{}\beta_{(g')^{-1}g''}
\]
giving the desired result multiplying by $(g')^{-1}$ in both sides. Observe that implicitly we are identifying $\frac{G\times G}{G}$ with $G$ via $[(g',g'')] \rightarrow (g')^{-1}g''$.
\end{itemize}
\qed

As stated, any symplectic transformation $f: T^*G \rightarrow T^*G$ produces a Lagrangian submanifold in $T^*G^-\times T^*G$, but also in $T^*(G\times G)$ using the {\it twist} mapping. The converse is also true, up to some regularity conditions. Given a Lagrangian submanifold ${\mathcal L}\subset T^*(G\times G)$ such that ${p_1^-}_{|{\mathcal L}}$ is a diffeomorphism, then there exists a mapping $\hat{\mathcal L}:T^*G \rightarrow T^*G$ making the diagram in Figure~\ref{fig:2} commutative,
\begin{figure}[H]
    \centering
    \[
\xymatrix{
& T^*(G\times G) \ar[dl]_{p_1^-} \ar[dr]^{p_2}&\\
T^*G \ar[rr]^{\hat{\mathcal L}} && T^*G
}
\]
    \caption{Illustration of the definition of the function $\hat{\mathcal L}$}
    \label{fig:2}
\end{figure}

It is natural to wonder whether the Lagrangian submanifold ${\mathcal L}^{red}\subset T^*G$ generates any sort of geometric transformation. As the  reader may have notice, this is the case. Using the theory of symplectic groupoids \cite{coste},  $T^*G$ can be seen as the symplectic groupoid integrating the Poisson manifold $\mathfrak{g}^*$.

\begin{theorem}[\cite{FLMV}]\label{poisson_mapping} Given a Lagrangian submanifold ${\mathcal L}$ in $T^*G$ such that
  $(J_L)_{|{\mathcal L}}:{\mathcal L}\rightarrow \mathfrak{g}^*$ is a (local)
  diffeomorphism. Then, there exists a  (local) Poisson automorphism,
  that we denote by $\hat{{\mathcal L}}$,
  given by the following recipe: given $\mu_1\in\mathfrak{g}^*$ there is
  a unique $\alpha_g$ such that $J_L(\alpha_g)=\mu_1$. Then
\[
\hat{{\mathcal L}}(\mu_1)=J_R(\alpha_g).
\]
\end{theorem}
If ${\mathcal L}\subset T^*(G\times G)$ is an invariant Lagrangian submanifold and ${\mathcal L}^{\red}\subset T^*G$ is the corresponding ``reduced'' Lagrangian submanifold constructed following Theorem~\ref{equivariant_lagrangian},  then the transformations described above ``match'' making the  diagram in Figure~\ref{fig:my_label} commutative.
\begin{figure}[H]
    \centering
\[
\xymatrix{
T^*G\ar[rr]^{\hat{{\mathcal L}}} \ar[dd]^{J_R}&& T^*G \ar[dd]^{J_R}\\ \\
\mathfrak{g}^*\ar[rr]^{\widehat{{\mathcal L}^{red}}} && \mathfrak{g}^*
}
\]
    \caption{Illustration of how the equivariant and the reduced transformation match, making the diagram above commutative.}
    \label{fig:my_label}
\end{figure}

\begin{figure}
    \centering
\[
\xymatrix{
& \ar@{.>}[dd]^>>>>>>>{and \ Rec}{\mathcal L}\subset T^*(G\times G)\ar[ld]_{pr_1^-} \ar[rd]^{pr_2}\\
 T^*G \ar[dd]^{J_R} \ar[rr]^<<<<<<<<<<<{\hat{{\mathcal L}}}&&  T^*G \ar[dd]^{J_R}\\
&\ar@{.>}[uu]^<<<<<<<{Red.} {\mathcal L}^{red}\subset T^* G\ar[dl]_{ J_L}
\ar[rd]^{ J_R}\\
 \mathfrak{g}^*\ar[rr]^{\widehat{{\mathcal L}^{red}}}&&  \mathfrak{g}^*
}
\]
    \caption{The diagram above allows to keep track of the different structures presented in this paper. While $T^*(G\times G)$ is a symplectic groupoid integrating $T^*G$, $T^*G$ is a symplectic groupoid integrating the Poisson manifold $\mathfrak{g}^*$. The correspondence between Lagrangian bisections is precisely the correspondence described in Theorem~\ref{equivariant_lagrangian}.}
    \label{fig:my_label1}
\end{figure}
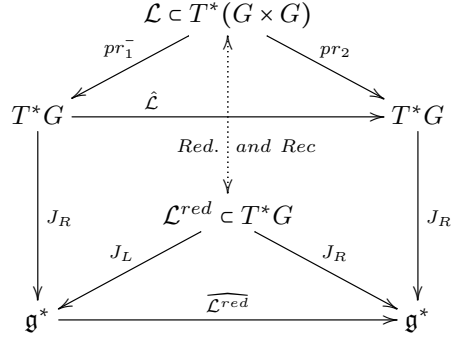

\paragraph{A Symplectic Groupoid Interpretation of our Constructions:} The diagram in Figure~\ref{fig:my_label1} presents a clean interpretation of our constructions in terms of symplectic groupoids.

\begin{proposition}[Reconstruction of Equivariant Mappings]
    Given a Lagrangian submanifold ${\mathcal L}^{red}\subset T^*G$ inducing a Poisson automorphism following Theorem~\ref{poisson_mapping}, then Algorithm~\ref{algorithm} allows us to obtain $\alpha^2_{g^2} = f(\alpha^1_{g^1})$, where $f$ is the corresponding equivariant mapping. That is, $twist(graph(f)) = {\mathcal L}$, where the reduction of ${\mathcal L}$ is precisely ${\mathcal L}^{\red}$.
\end{proposition}
\begin{algorithm}[H]
\caption{Reconstruction of Equivariant Mappings }\label{algorithm}
\begin{algorithmic}[1]

    \State Given $\alpha'_{g'} \in T^*G$ compute $\mu' = J_R(\alpha'_{g'})$
    \State Take the only element in $ \alpha_g\in {\mathcal L}^{red}$ such that $J_L(\alpha_g) = \mu'$
    \State Compute $\mu'' = J_R(\alpha_g)$
    \State Compute $g'' = g'\cdot g $
    \State Compute $\alpha''_{g''}$ by solving the linear equation $J_R(\alpha''_{g''}) = \mu''$
    \end{algorithmic}
\end{algorithm}
\proof
Keeping track of the correspondence between a Lagrangian submanifold and the reduced Lagrangian submanifold directly yields the result.
\qed

\begin{proposition} The mapping constructed by Algorithm~\ref{algorithm} is $G$-equivariant and conserves the momentum mapping $J_L$.
  \end{proposition}

\proof 
The proof is a straightforward but lengthy computation using the definitions of $J_R$ and $J_L$ and the correspondence presented in Theorem~\ref{equivariant_lagrangian}.
\qed



  \section{Applications to Simulation and Learning of Systems}

The constructions above are theoretically sound but, most importantly, they allow us to tailor geometric integrators and learning mechanisms that conserve the underlying geometry, symplectic and Poisson, while preserving symmetry. The main idea is to use well-known techniques in symplectic geometry to parameterize the Lagrangian submanifolds of interest, which are the ones in $T^*G$. For instance, we will resort to the technique of generating functions (\cite{marsden3}), but any other technique can be used (like Morse families). The Lagrangian submanifolds in $T^*G$  are in one-to-one correspondence with $G$-equivariant transformations which can be easily computed through the reconstruction procedure depicted in Algorithm~\ref{algorithm}, yielding the desired mappings that can be used to simulate or learn dynamical systems.
\subsection{Simulating Hamiltonian Systems with Symmetry}

In this section we deal with the problem introduced in Section~\ref{sec:simulation_of_systems_with_symmetries}. That is, we are given a $G$-invariant Hamiltonian $H: T^*G\rightarrow \mathbb{R}$ and we seek to construct both a $G$-equivariant symplectic integrator that approximates $\phi^H_t$ and a Poisson integrator  that approximates $\phi^{H^\red}_t$. Moreover, both integrators must satisfy a commutativity condition analogous to the one depicted in Figure~\ref{fig:1}. To achieve this objective, we make use of the constructions presented in the previous section, along with other geometric insights pertaining generating functions from~\cite{FLMV}.
\vspace*{0.25cm}

{\it $\bullet$ Step 1: Reduce the Problem.}  Since $\phi^H_t$ is symplectic and $G$-equivariant, due to the correspondence explained in Theorem~\ref{equivariant_lagrangian}, we can look for a Lagrangian submanifold, say ${\mathcal L}^{\red}$, in $T^*G$ that corresponds to the desired transformation. 
\vspace*{0.25cm}

{\it  $\bullet$ Step 2: Solve the Reduced Problem.} In order to look for  ${\mathcal L}^{\red}$, one should notice that ${\mathcal L}^{\red}$ is the Lagrangian submanifold in $T^*G$ that generates the flow $\phi^{H^{\red}_t}$. One way to accomplish this is through the theory of Hamilton-Jacobi developed in Section $3.6$ of \cite{FLMV}, which primarily involves parameterizing Lagrangian submanifolds near the identity by employing a specific class of generating functions, namely those of the form
      \[
{\mathcal L}^{\red}_S= \{(\displaystyle\frac{\partial S}{\partial p}(p), p)\in T^*G\}.
        \]
Allowing time-dependent generating functions, we can look for functions satisfying the Poisson Hamilton-Jacobi equation, which reads
\[
\displaystyle\frac{\partial S}{\partial t}(t,p) + H(J_R(\frac{\partial S}{\partial p}(t,p),p)) = 0.
\]
Even when the last expression is hard to solve, methods to approximate it exist. Here we follow the approach in~\cite{FLMV} but any other means can be used, see Section~\ref{sec:Conclusions and Future Work}. Notice that the approximation of the solution of the Hamilton-Jacobi theory is independent of the other constructions and therefore any other approximation strategy can be combined with our setting.

\vspace*{0.25cm}

{\it $\bullet$ Step 3: Reconstruct the Original Problem (Un-reduce).} Once the Lagrangian submanifold ${\mathcal L}^{\red}$ has been approximated through a generating function, then ``un-reduce'' using the reconstruction procedure presented in Algorithm~\ref{algorithm}. This  provides a symplectic, $G$-equivariant, momentum preserving approximation to $\phi^H_{t}$.

\subsubsection{Example: Rigid Body}

This section is devoted to illustrate the outcome of the integrators designed with our procedure. We show the outcome of un-reducing the Poisson integrators generated by the Hamilton-Jacobi theory from~\cite{FLMV}. To follow the tradition, we apply our findings to the rigid body (see \cite{marsden3}), with reduced Hamiltonian
\[
H^{\red}(x,y,z) = \displaystyle\frac{1}{2}\left(\frac{x^2}{1.5} + \frac{y^2}{2} + \frac{z^2}{2.5}\right).
\]
We approximate the Hamilton-Jacobi equation following~\cite{FLMV}, through a Taylor's expansion of the equation in the $t$-variable and truncating the resulting recurrence equation to certain order, which is made explicit in each case. We obtain truncations of the reduced Hamilton-Jacobi equation of order $3$, $5$ and $7$, and  use Algorithm~\ref{algorithm} to simulate the dynamics of the left-invariant Hamiltonian $H = H^{\red}\circ J_R$.
\begin{figure}
    \centering
    \includegraphics[scale = 0.4]{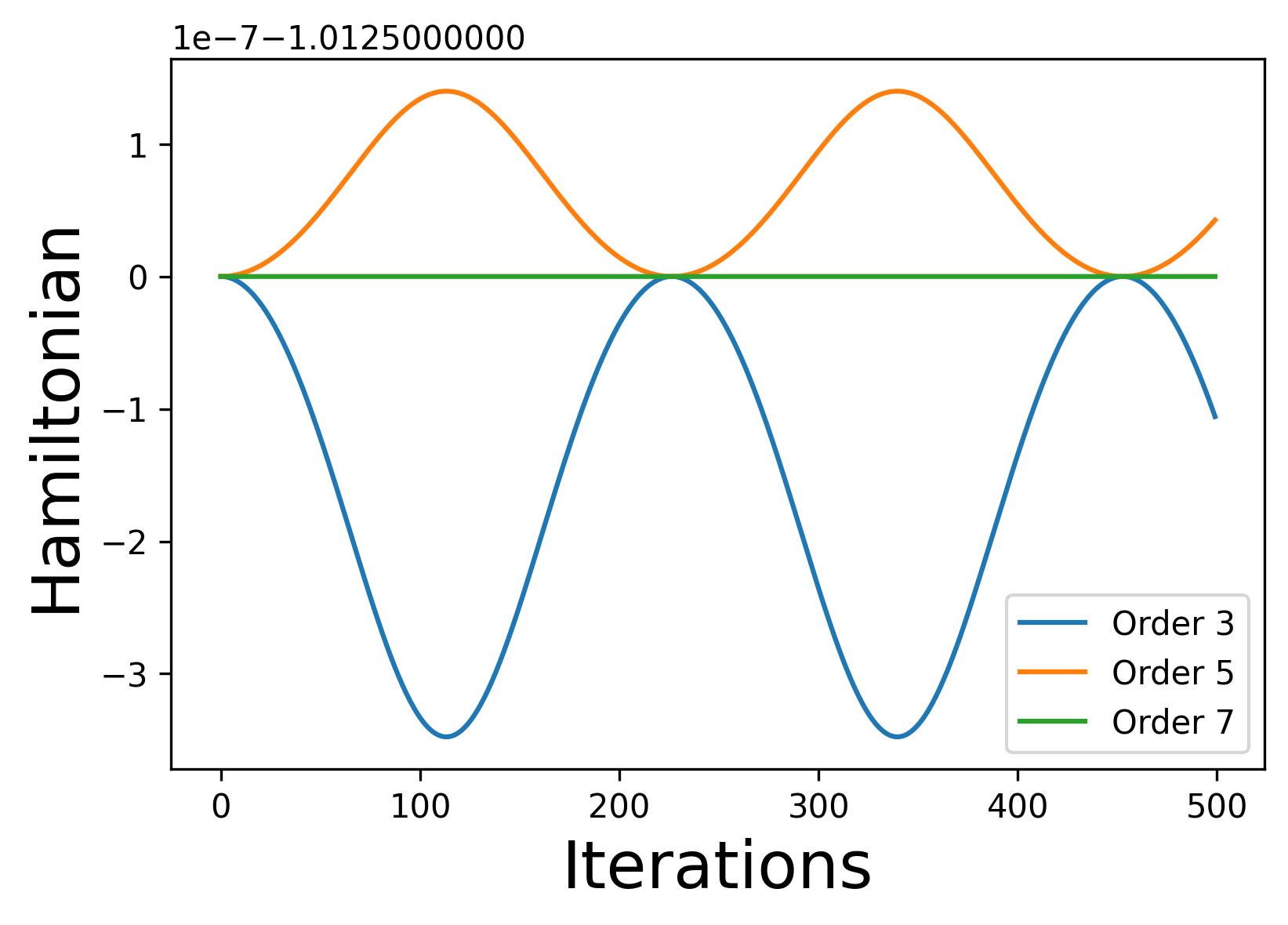}
    \includegraphics[scale = 0.4]{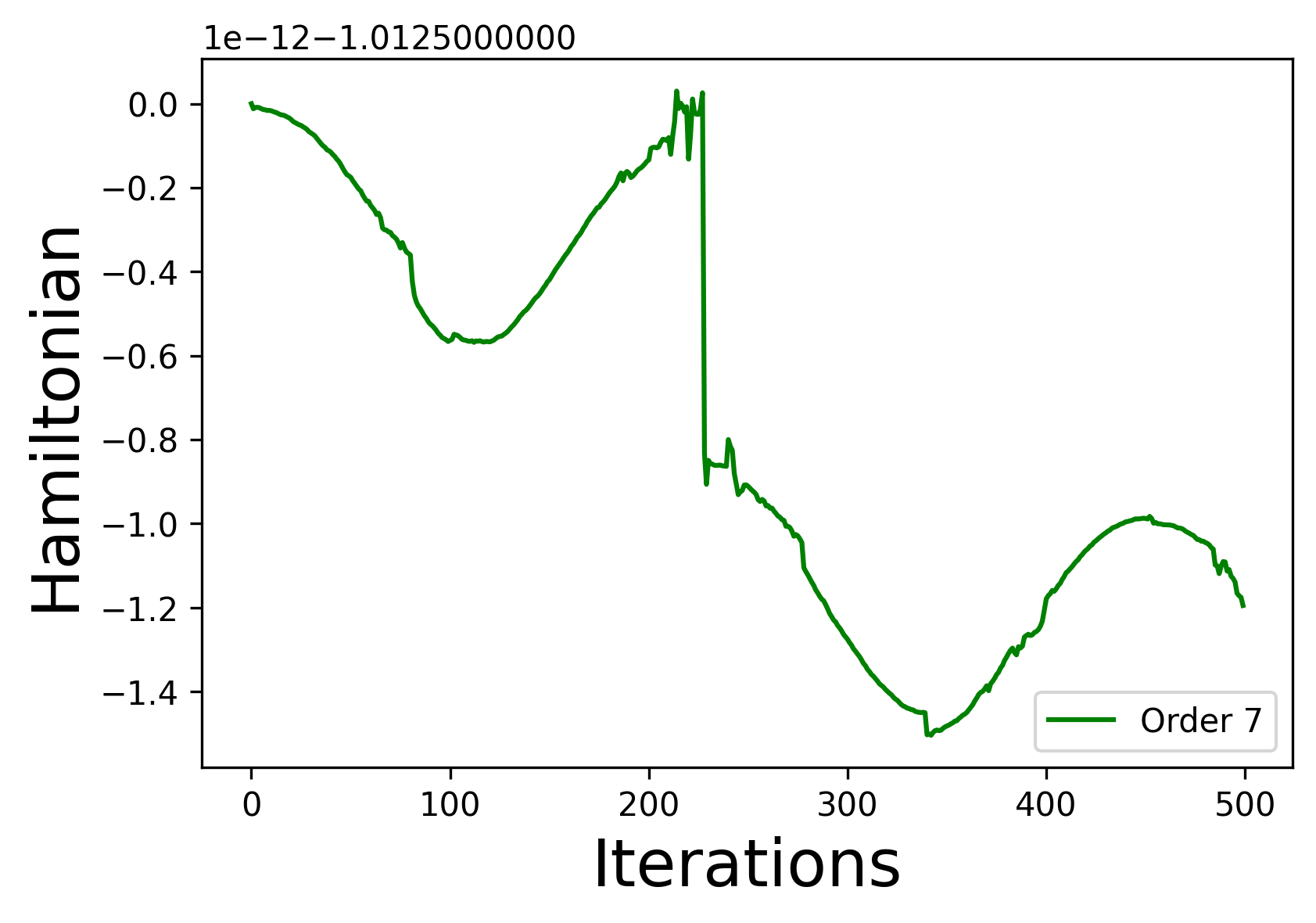}
    \includegraphics[scale = 0.4]{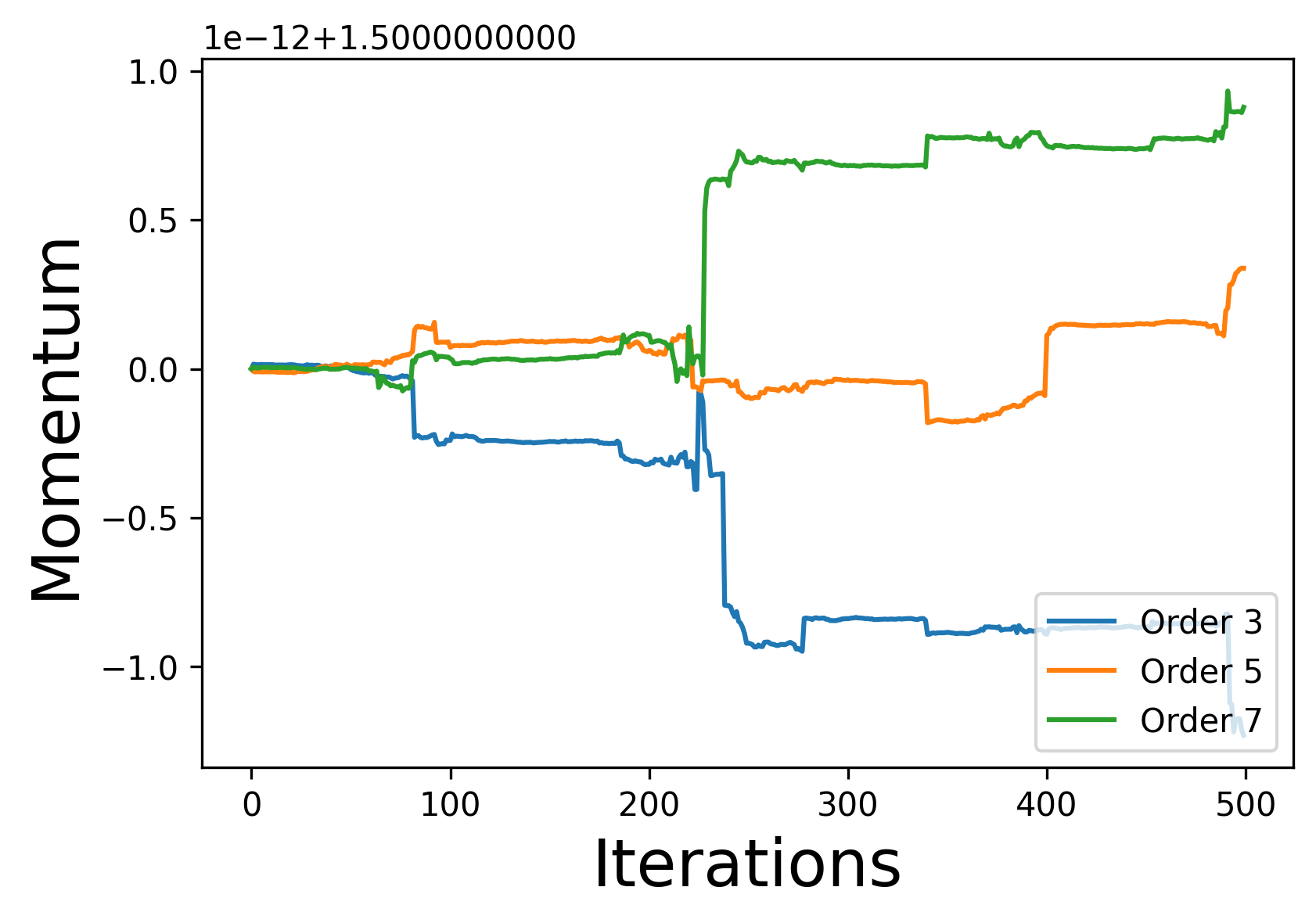}
    \includegraphics[scale = 0.4]{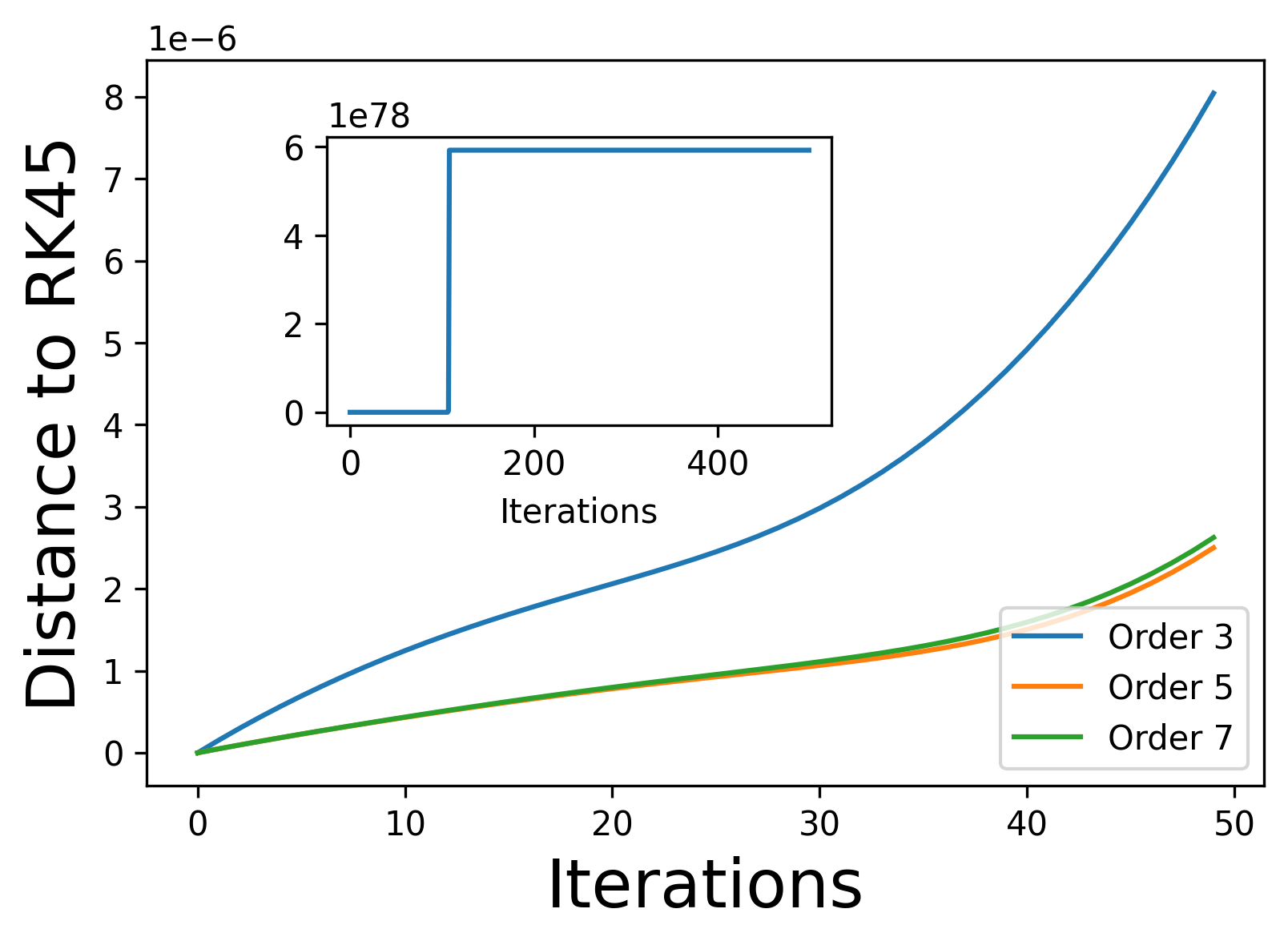}
    \caption{
    {\it Top-Left:} Evolution of the different order integrators. We observe how higher-order integrators tend to conserve better the system's Hamiltonian. Nonetheless, all orders show oscillations of the Hamiltonian function around the initial value.
    Comparison of the evolution of the momentum mapping for our integrator and RK45. 
    {\it Top-Right:} Evolution of the Hamiltonian for order $7$ integrator. We observe a very good energy preservation.
    {\it Bottom-Left:} We see the evolution of the three components of the momentum mappings we our integrator is used. It comes at no surprise that the momentum is conserved up to numerical error, by construction. On the right we see how RK45 is oscillating with the oscillations becoming quite rough at some points.
    {\it Bottom-Right:} Comparison with variable step-size Runge-Kutta 4-5. We observe how the difference is small for the first steps of the iteration but eventually the RK45 integrator blows up while the geometric integrators are able to simulate the Hamiltonian dynamics.
    }
    \label{fig:my_label}
\end{figure}

\subsection{Learning Hamiltonian Systems with Symmetry}
        In this section we consider the problem of learning Hamiltonian systems with symmetry described in Section~\ref{sec:learning_systems_with_symmetry}. Given the data set of pairs $D = \{(\alpha'_{g'},\alpha''_{g''})_i,\ i \in I\}$ corresponding to a Hamiltonian system evolving on a Lie group where the momentum mapping is preserved,  our main goal is to learn the $G$-equivariant transformation $\phi^H_{\Delta t}$ that generated the data, that is, the one satisfying $\phi^H_{\Delta t}(\alpha'_{g'}) = \alpha''_{g''}$. We follow a strategy parallel to the one in the previous section.
\vspace*{0.25cm}
        
  {\it $\bullet$ Step 1: Reduce the Problem.} Our aim is to look for a $G$-invariant Lagrangian submanifold that passes by the points in the data set $D$. Nonetheless, since we do not have a means to directly parameterize these Lagrangian submanifolds, we create the reduced data set $D^\red\subset T^*G$ as follows
      \[
\textrm{If } (\alpha'_{g'},\alpha''_{g''})\in D \textrm{ then } (g')^{-1}\cdot{}\alpha''_{g''}\in D^{red}.
        \]
    This construction is motivated by Theorem~\ref{equivariant_lagrangian}.
    
     {\it $\bullet$ Step 2: Solve the Reduced Problem.} The next point consists of learning a Lagrangian submanifold $ {\mathcal L}\subset T^*G$ such that $D^{\red}\subset {\mathcal L}$. This contrasts with other recent approaches in the literature where generating functions are also used. The main question is how to design a learning procedure of Lagrangian submanifolds that is valid in our context. Following  \cite{FLMV}, we parameterize all Lagrangian submanifolds close the the identify by generating functions $S(p)$ and through the expression
        \[
{\mathcal L}^{\red}_S= \{(\displaystyle\frac{\partial S}{\partial p}(p), p)\in T^*G\}.
        \]
      Now that we have a reduced data-set $D^{\red}$ and a means to generate all relevant Lagrangian submanifolds, we train a feed-forward neural network that parameterizes all the functions $S(p)$, that is $S(p;W)$ where $W$ are the weights of the neural network. We use the mean squared error (MSE) to guide the training process, which aims at solving the optimization problem
        \begin{align}\label{optimization-equation}
\min\limits_{W} & \sum_{i \in I}(g^i - \displaystyle\frac{\partial S}{\partial p}(\alpha^i;W))^2
          \end{align}
where the expression above should be understood in canonical coordinates around the identity element of the Lie group $G$ and $(g^i, \alpha^i)\equiv \alpha_{g^i}\in D^{\red}$.

\begin{remark}[Alternative Optimization Problem]  Since we are solving an optimization problem on a Lie group and $g^i$ is near the identity of the Lie group, we can use the exponential or a similar retraction map to solve the alternative problem:
\begin{align}\label{optimization-equation-1}
\min\limits_{W} & \sum_{i \in I}\left(\hbox{exp}^{-1}\left[(g^i)^{-1}\displaystyle\frac{\partial S}{\partial p}(\alpha^i;W)\right]\right)^2
          \end{align}
 \end{remark}

\vspace*{0.25cm}

 {\it $\bullet$ Step 3: Reconstruct the Original Problem (Un-reduction).} The previous step allows us to find a Lagrangian submanifold that approximates the reduced data-set $D^{\red}$. We are interested in learning the original mapping, $\phi^H_{\Delta t}$. In order to succeed we can apply the reconstruction procedure introduce in Algorithm~\ref{algorithm}.

\begin{remark}
Alternatively, given the data set $D$  we define  the elements
$\mu'=J_R(\alpha'_{g'})$
and $\mu''=J_R(\alpha''_{g''})$ 
in ${\mathfrak g}^*$. With this reduced data set $D^{\red}$ we  try to learn the reduced Hamiltonian $H^{\red}: {\mathfrak g}^*\rightarrow {\mathbb R}$ using Lie-Poisson integrators \cite{bourabee04,david}
$$
\min\limits_{W}  \sum_{i \in I}\left(\xi_i-\frac{\partial H^{\red}}{\partial \mu}\left(\frac{1}{\Delta t}(d_L\hbox{exp}_{\Delta t \xi_i})^*\mu''; W\right)\right)^2
$$      
where $W$ are the weights of the neural network for   the reduced Hamiltonian,  $\xi_i=\frac{1}{\Delta t}\hbox{exp}^{-1}(g^i)$ and $d_L\hbox{exp}_{\xi}: {\mathfrak g}\rightarrow \mathfrak{g}$ is the left trivialized tangent map \cite{bourabee04}.   
 After the optimization process we obtain a Hamiltonian $H^{\red}_{W^*}: {\mathfrak g}^*\rightarrow {\mathbb R}$ and the un-reduced one  $H_{W^*}(\alpha_g)=H^{\red}_{W^*}(g^{-1}\cdot \alpha_g)$ for all $g\in G$.
\end{remark}

\subsubsection{Example: The Rigid Body}

\begin{figure}[H]
    \centering
    \includegraphics[scale = 0.4]{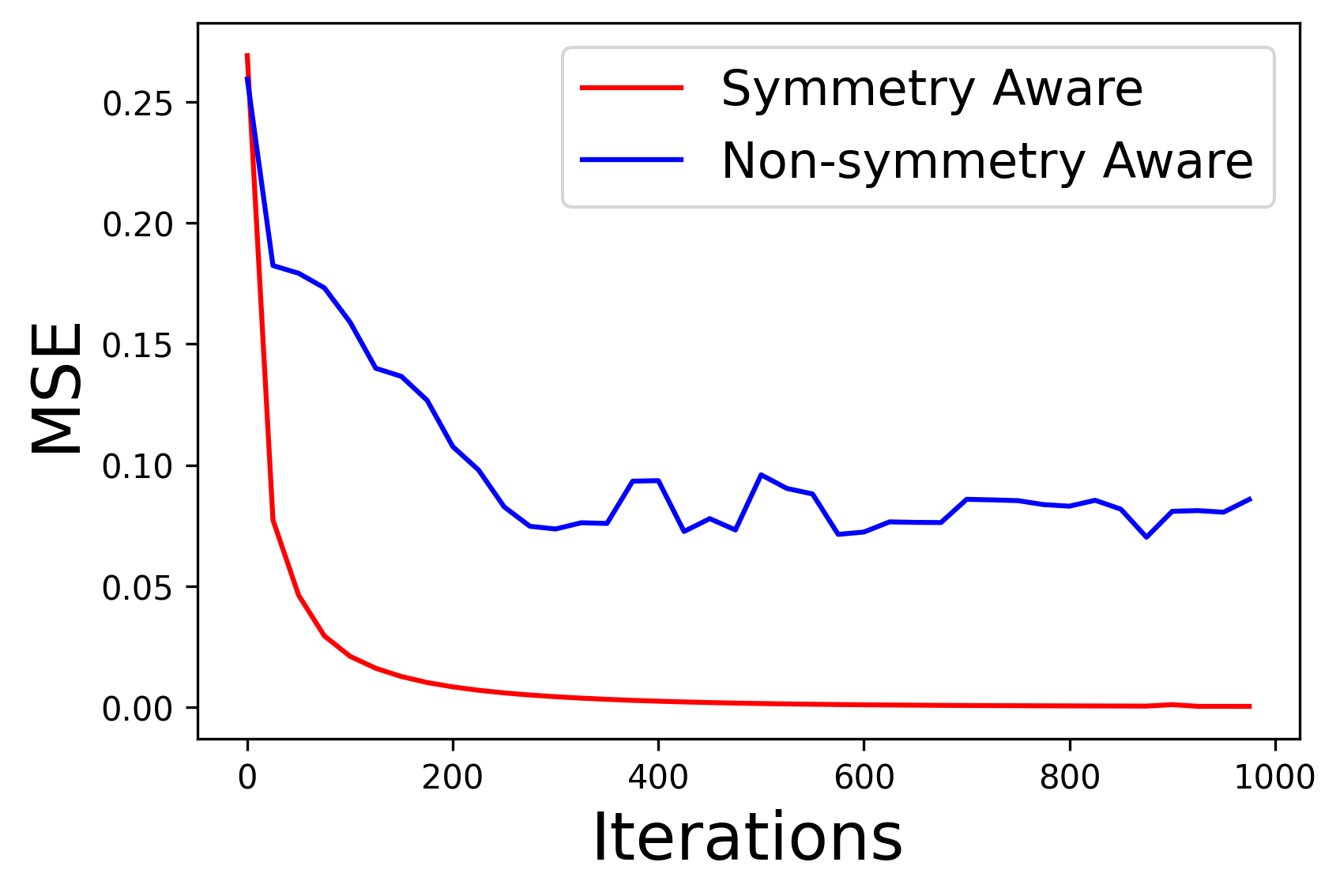}
    \includegraphics[scale = 0.4]{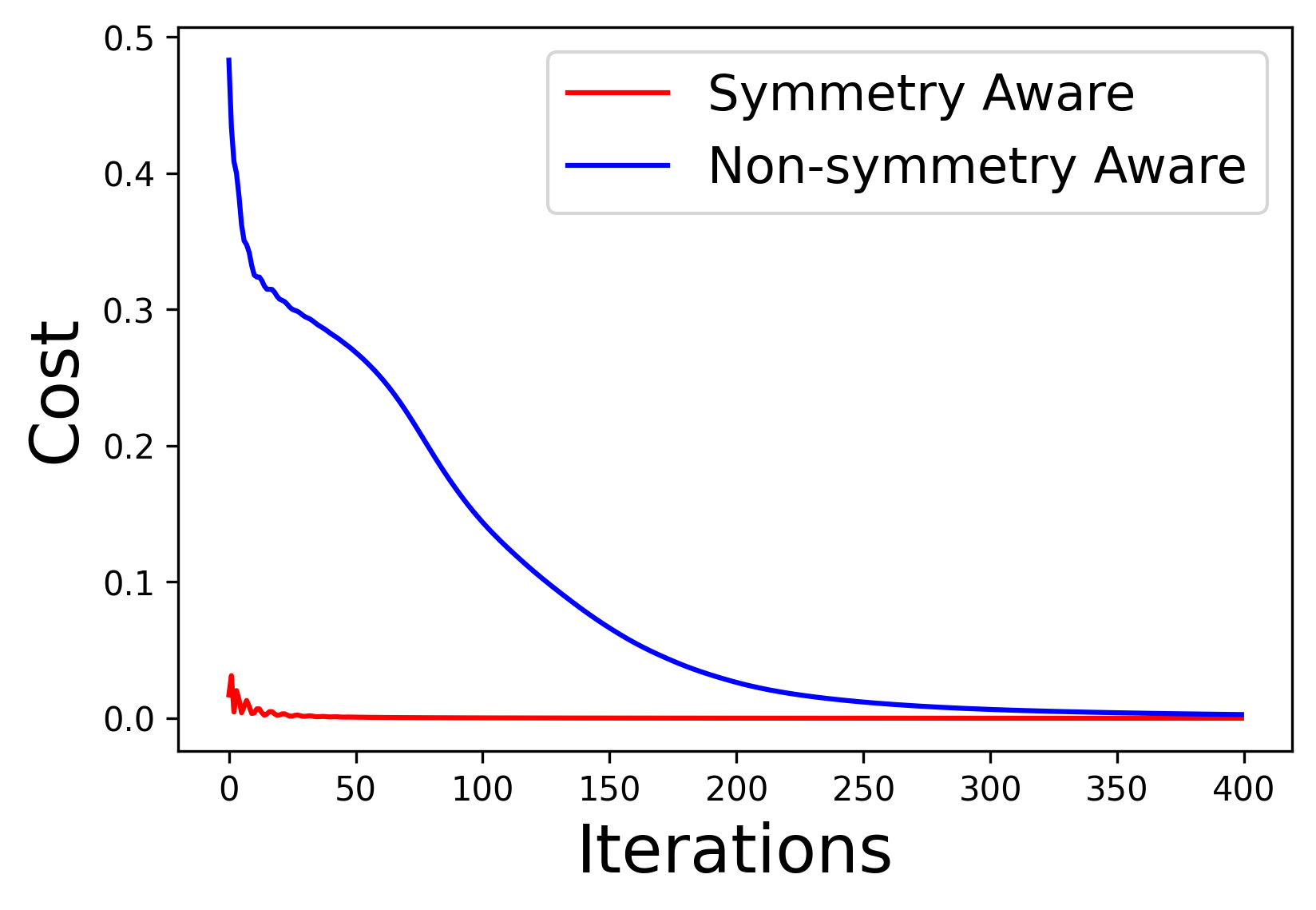}
    \includegraphics[scale = 0.4]{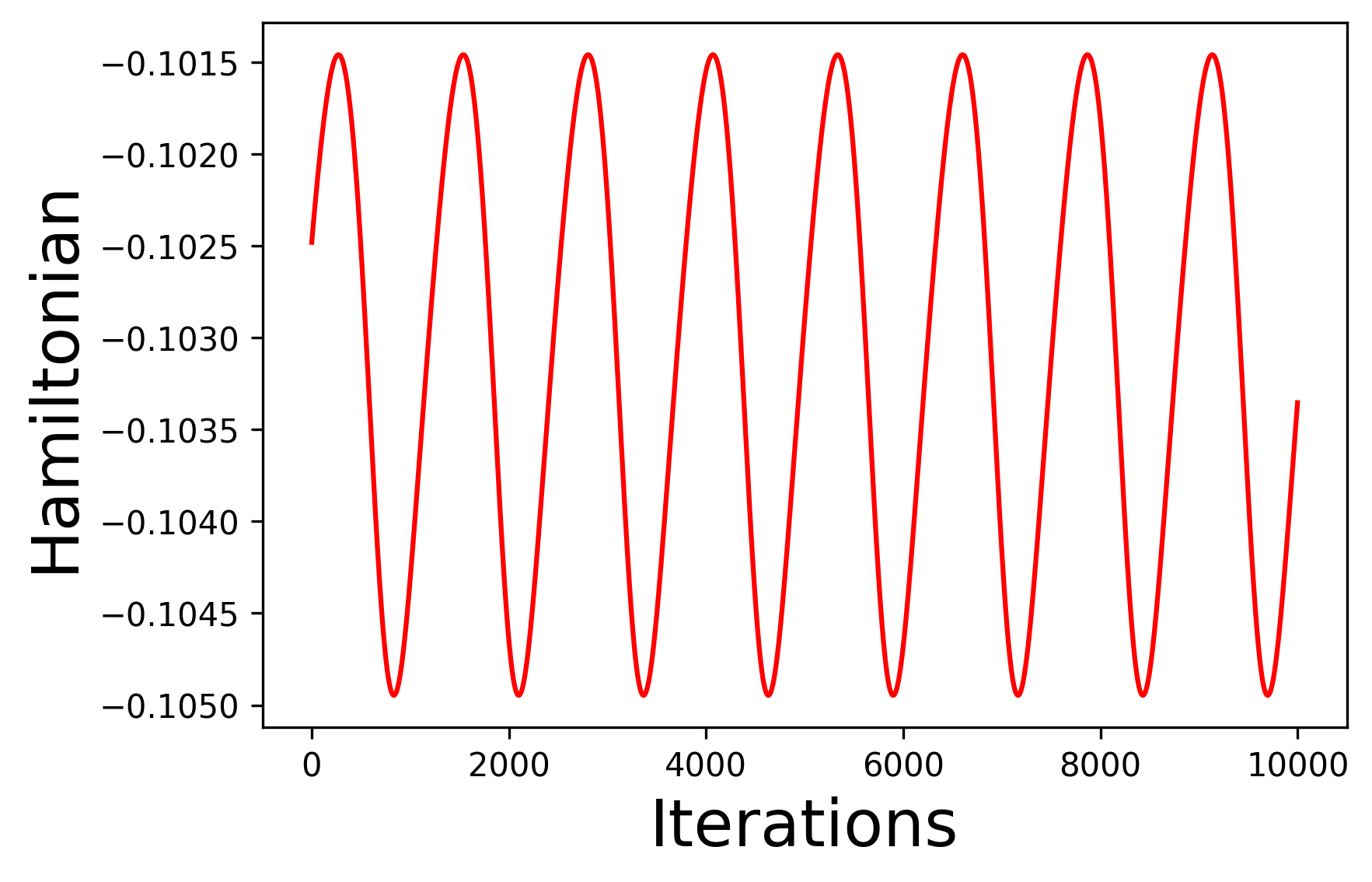}
    \includegraphics[scale = 0.4]{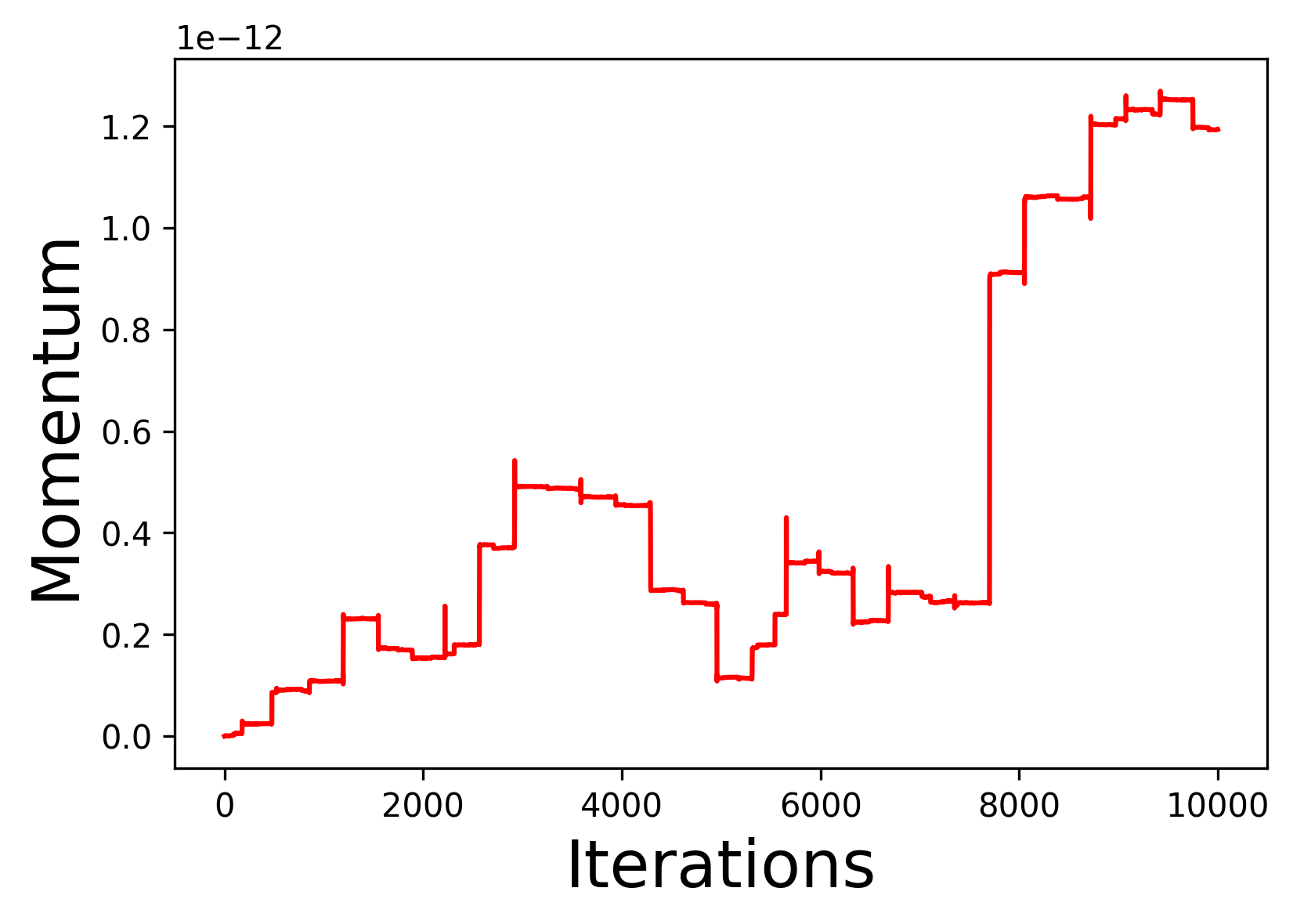}
    \caption{{\it Top-Left} and {\it Top-Right:} We observe that training the reduced model seems easier than the un-reduced one. The proposed model usually starts closer to the solution and the evolution of the MSE is faster. This results in better prediction over new data. 
    {\it Bottom-Left:} We plot here the evolution of a trajectory obtained using our trained model. Not surprisingly, we obtain an evolution similar to the one obtained with symplectic integrators.
    {\it Bottom-Right:} As expected, the momentum mappings is conserved up to round-off error.
    }
    \label{fig:my_label3}
\end{figure}

We learn here the dynamics of the rigid body with the same parameters as described in Section~\ref{sec:simulation_of_systems_with_symmetries}. We generate a data set using the integrators developed in this paper, which ensure we obtain samples from an equivariant transformation. We analyze the inclusion of symmetry in the training process, comparing the procedure described here with the use of mere generating function following~\cite{chen-tao}.  We train a feed-forward neural network with just two layers of $500$ neurons using $500$ points in $T^*G$ using the Cayley mapping, see~\cite{Ferraro_2015}. The entries of the points are uniformly randomly generated between $0$ and $6$. We train our models for $1000$ steps using the Adam optimizer with step-size $0.001$. The outcome of the process is collected in Figure~\ref{fig:my_label3}, where {\it Symmetry Aware} denotes the model obtained following the constructions presented here and {\it Non-symmetry aware} refers to the model without symmetry following~\cite{chen-tao}.

We also assess the performance of the symmetric and non-symmetric models for different situations collected in Table~\ref{table1}.
\begin{table}[H]
\begin{tabular}{|c|c|c|c|c|c|}
\hline
          \textbf{Step-size}   &  \textbf{N. Samples}    & \textbf{Type} &  \textbf{MSE} & \textbf{Max. Error} &  \textbf{Gradient}\\ \hline \midrule
          \multirow{2}{*}{0.05} & \multirow{2}{*}{100} & Symmetric  & 0.0024 & 1.0625 & 2.22e$^{-5}$ \\ \cline{3-6} 
                  &                   &  Non-symmetric & 0.0202& 1.4479 &0.0001 \\ \hline \midrule        
\multirow{2}{*}{0.05} & \multirow{2}{*}{500} & Symmetric  & 0.0012 & 0.7134 &  0.0720 \\ \cline{3-6} 
                  &                   &  Non-symmetric & 0.0063 & 1.1210 & 0.0176 \\ \hline \midrule
\multirow{2}{*}{0.1} & \multirow{2}{*}{500} & Symmetric  &  0.0042 &  1.0496  & 3.19$e^{-5}$ \\ \cline{3-6} 
                  &                   &  Non-symmetric & 0.0429 & 2.0936 &  0.0448 \\ \hline \midrule
\multirow{2}{*}{0.1} & \multirow{2}{*}{1500} & Symmetric  &  0.0002 & 0.4823 & 6.23$e^{-6}$ \\ \cline{3-6} 
                  &                   &  Non-symmetric & 0.0482 & 4.65 &  $0.0176$ \\ \hline 

\end{tabular}
\caption{ Outcome of training the two neural networks in different situations, where the training data is simulated using different step-sizes and changing the size of the data set. We observe that the symmetric model always outperforms the non-symmetric one as expected. The column MSE reflects the mean squared error of testing the two models on $100$ points simulated like the training data set, not seen by the models before. The Max. Error column collects the maximum disagreement between the ground truth and the training model on the testing data. The Gradient column shows the norm of the gradient at the last iteration. For the same number of steps, we observe that the symmetric model tends to obtain a lower value of the gradient, reflecting that it is easier to train.}\label{table1}
\end{table}

\begin{figure}[H]
    \centering
    \includegraphics[scale = 0.4]{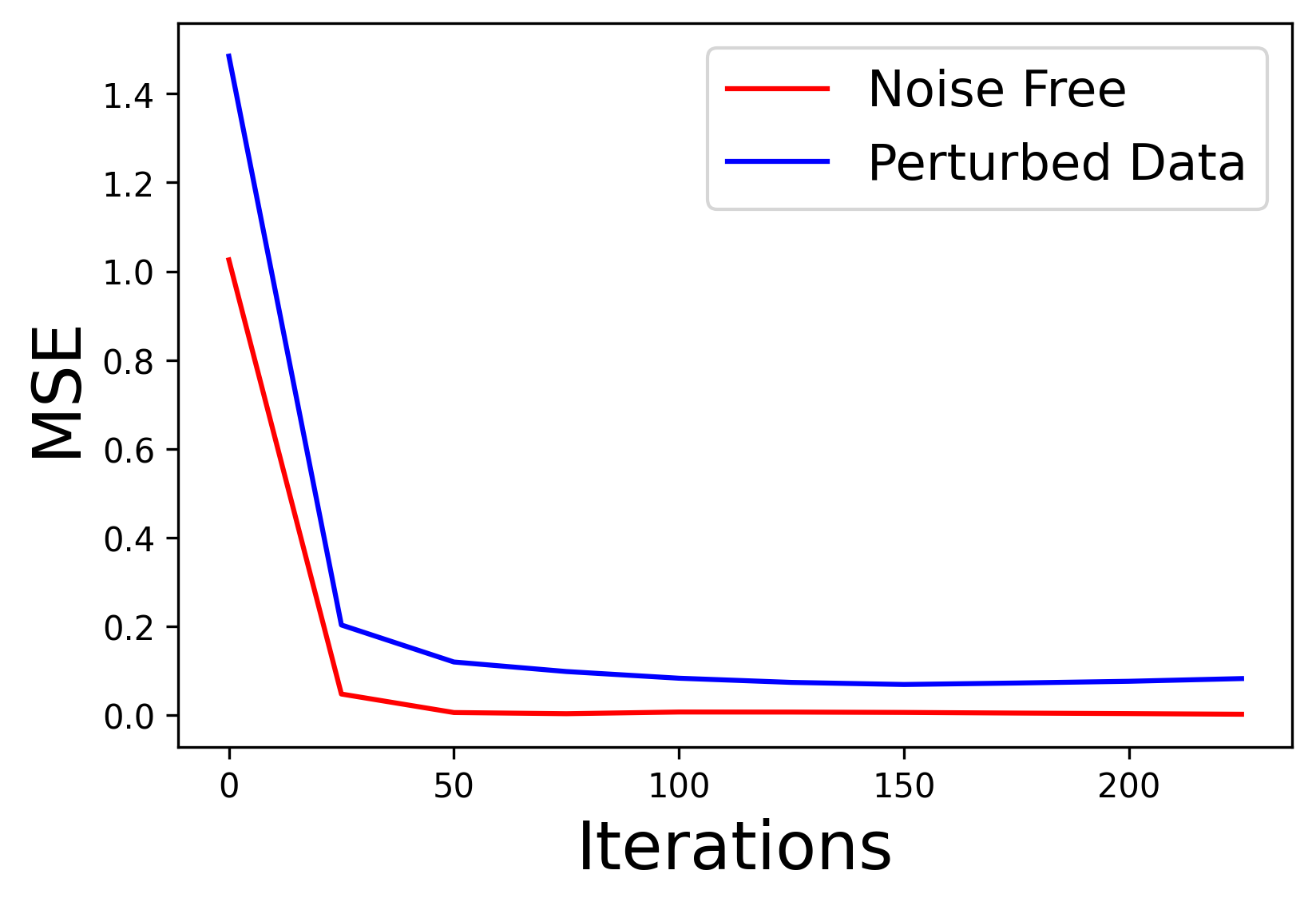}  
    \includegraphics[scale = 0.4]{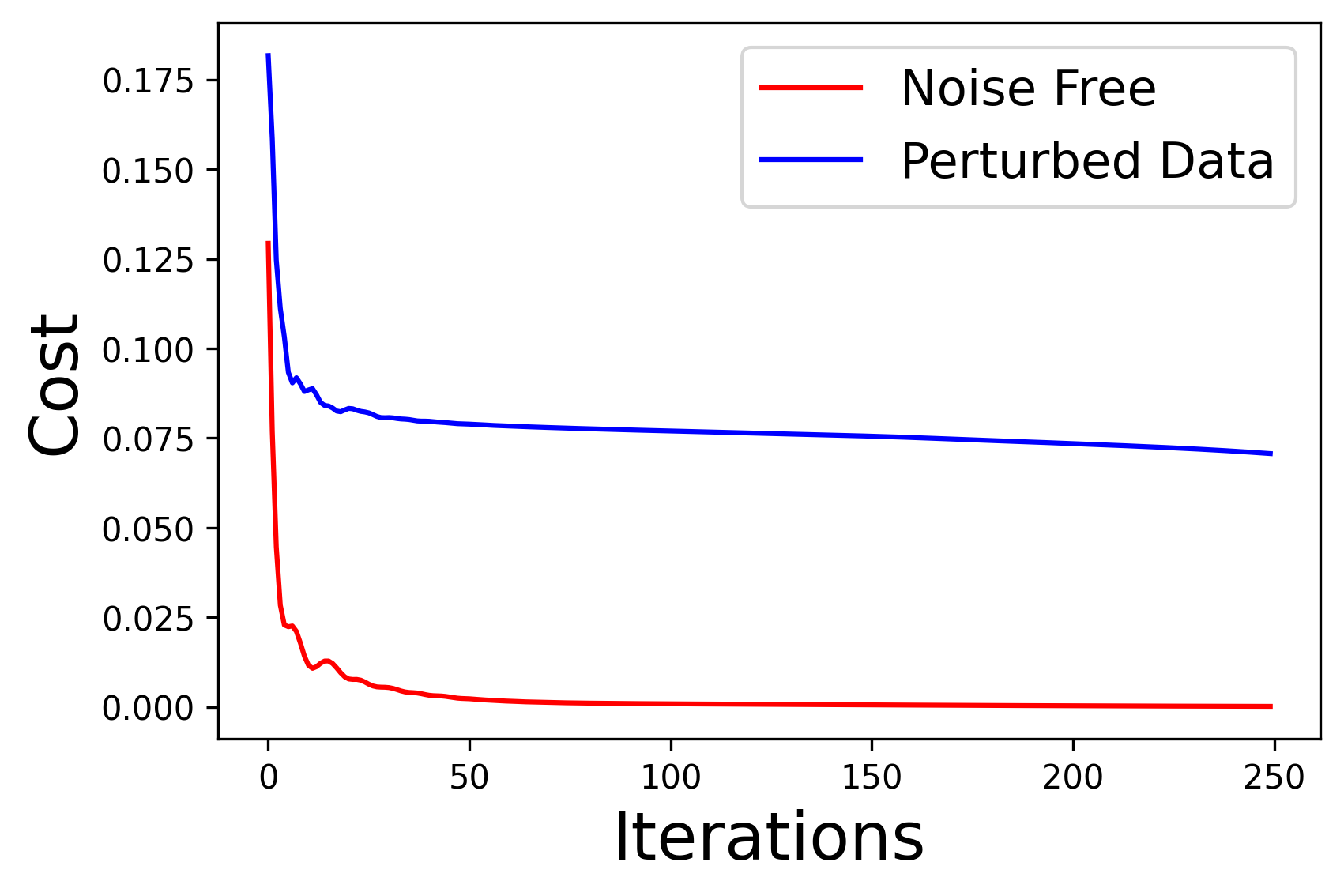}
    \caption{{\it Left:} Evolution of MSE on testing data thought the training process. We observe how, in spite of the noise, the model trained with perturbed data is able to achieve good prediction on unseen data. {\it Right:} Evolution of the cost (MSE) throughout the training process. While the model with noise-free data is able to adjust nicely to the data, the model with perturbations find the $G$-equivariant symplectic transformation that is closer to the data (neither symmetric, neither symplectic), which explains why the model is unable to decrease the cost further.}
    \label{fig:my_label4}
\end{figure}

We have also tested the robustness against noise of the proposed approach, with promising results. In Figure~\ref{fig:my_label4} we show the outcome of comparing two models, with and without noise, using the procedure introduced in this paper. One model is trained using noise-free data consisting of $1000$ samples of the rigid body obtained using a symplectic integrator with step-size $0.15$, while the other is trained using additive perturbations sampled from a Gaussian distribution with mean $0$ and variance $0.05$. We train a feed-forward neural network with two layers of $500$ neurons like before and observe the effect of perturbing the data. We  train these models for $250$ iterations using Adam and show the outcome below.

To finalize this section, we provide a theoretical result concerning the behaviour of the error under the reduction process for the rigid body. Since we are dealing with manifolds and there is no straightforward notion of additive error, we resort to the ambient space and its inner product, as the space of $3\times 3$ matrices can be identified with $\mathbb{R}^9$. Under this identification, the standard scalar product becomes the Frobenius inner product of matrices. This permits the construction of an embedding $T^*SO(3) \rightarrow T^*\mathbb{R}^9 \equiv \mathbb{R}^9 \times \mathbb{R}^9$, just considering the linear extension of mappings that associates the zero mapping to the normal of $TSO(3)$ at each point. Thus, every point $\alpha_g$ can be identified with a pair $(g,\alpha)\in \mathbb{R}^9\times \mathbb{R}^9$ and the Lie group action translates in the obvious way.

\begin{proposition}[Error Transformation Under Reduction for the Rigid Body]
    Under the previous identifications, assuming the pair $(g',\alpha')$  $(g'',\alpha'')$ are in the data set, if we perturb them to be $(\tilde{g}',\tilde{\alpha}') = (g' + \epsilon'_1,\alpha' + \epsilon'_2)$ and $(\tilde{g}'',\tilde{\alpha}'') = (g'' + \epsilon''_1,\alpha'' + \epsilon''_2)$ such that $\tilde{g}',\ \tilde{g}''\in SO(3)$ then
\[
\norm{(g')^{-1}g''- (\tilde{g}')^{-1}\tilde{g}''} = \mathcal{O}(\norm{\epsilon'_1} + \norm{\epsilon'_2} + \norm{\epsilon'_1}\norm{\epsilon'_2})
\]
and
\[
\norm{(g')^{-1}\cdot {}\alpha'' - (\tilde{g}')^{-1}\cdot{}\tilde{\alpha}''} = \mathcal{O}(\norm{\tilde{\alpha}''}\norm{\epsilon'_2} + \norm{\epsilon''_2} + \norm{\epsilon''_2}\norm{\epsilon'_2})
\]
where $g\cdot{} \alpha$ just refers to the Lie group action understood after the identification  $T^*SO(3) \rightarrow T^*\mathbb{R}^9 \equiv \mathbb{R}^9 \times \mathbb{R}^9$. Since in finite dimensions all norms are equivalent we do not make any specific choice of the used norm.
\end{proposition}
\proof
We only prove the first inequality and the second one follows applying similar arguments. We have
\begin{align*}
&(g')^{-1}g''- (\tilde{g}')^{-1}\tilde{g}'' = (g')^{-1}g'' - (g' + \epsilon'_1)^T(g''  + \epsilon''_1),  \\    
&= (g')^Tg'' - (({g'})^T + ({\epsilon'_1})^T)(g''  + \epsilon''_1),
\end{align*}
and using that $(g' + \epsilon'_1)\in SO(3)$
\begin{align*}
&(g')^{-1}g'' - (g' + \epsilon'_1)^{-1}(g''  + \epsilon''_1) = (g')^{-1}g'' - (g' + \epsilon'_1)^{-1}(g''  + \epsilon^2_1) \\
&= (g')^T \epsilon''_1 + (\epsilon'_1)^T g'' + (\epsilon'_1)^T g'',
\end{align*}
which after taking norms and noting that $\norm{g'}$ and $\norm{g''}$ are bounded for being orthogonal matrices gives the result.
\qed

\begin{remark}[Practical Justification] The previous result just manifests that when error is present, the reduced data set is modified linearly in the error as long as the momentum is bounded. Since in applications this is a reasonable assumption, the result provides a theoretical framework that justifies our approach. A similar result  can be obtained for other matrix Lie groups, although the equivalent statement for general Lie groups would require more involved tools due to the lack of normed vector space structure.
\end{remark}
    
\subsection{Learning Poisson Dynamical Systems}
A system with symmetry contains redundant information, and therefore, the essential information can be considered to reside in a quotient space. This quotient space represents the reduced system by removing the repetitive information. If the original Hamiltonian system possesses a symplectic structure, then the quotient space will exhibit a well-established structure known as Poisson structure.

In Section~\ref{sec:learning_systems_with_symmetry}, we outlined our approach for indirectly learning Poisson transformations by utilizing the reduced data set $D^\red$. Thus, a  reasonable question is still open: 
\begin{center}
    {\it Can we learn Poisson dynamical systems respecting the underlying geometry?}
\end{center}
To be more explicit, we assume that a data set is given $D = \{(\mu',\mu'')_i,\ i \in I\} \subset \mathfrak{g}^*$ generated by an unknown Hamiltonian $H:\mathfrak{g}^*\rightarrow \mathbb{R}$ with time step $\Delta t$, then we would like to learn $\phi^H_{\Delta t}$. Therefore we need to look for a Lagrangian submanifold ${\mathcal L} \subset T^*G$ such that there exist points $(g_i,p_i)\in L$ such that
\[
J_R((g_i,p_i)) = (\mu')_i , \quad J_L((g_i,p_i)) = (\mu'')_i.
\]
The main problem is how to compute this points $(g_i,p_i)$ that before where given by the reduced data set. We propose to generate the candidate Lagrangian submanifolds like before, using generating functions and to obtain these points through the equation
\[
g_i = \displaystyle\frac{\partial S}{\partial p}(p_i:W)
\]
where $p_i$ are also parameters to be learnt. That is, they also form part of our neural network.

\begin{figure}
    \centering
    \includegraphics[scale = 0.3]{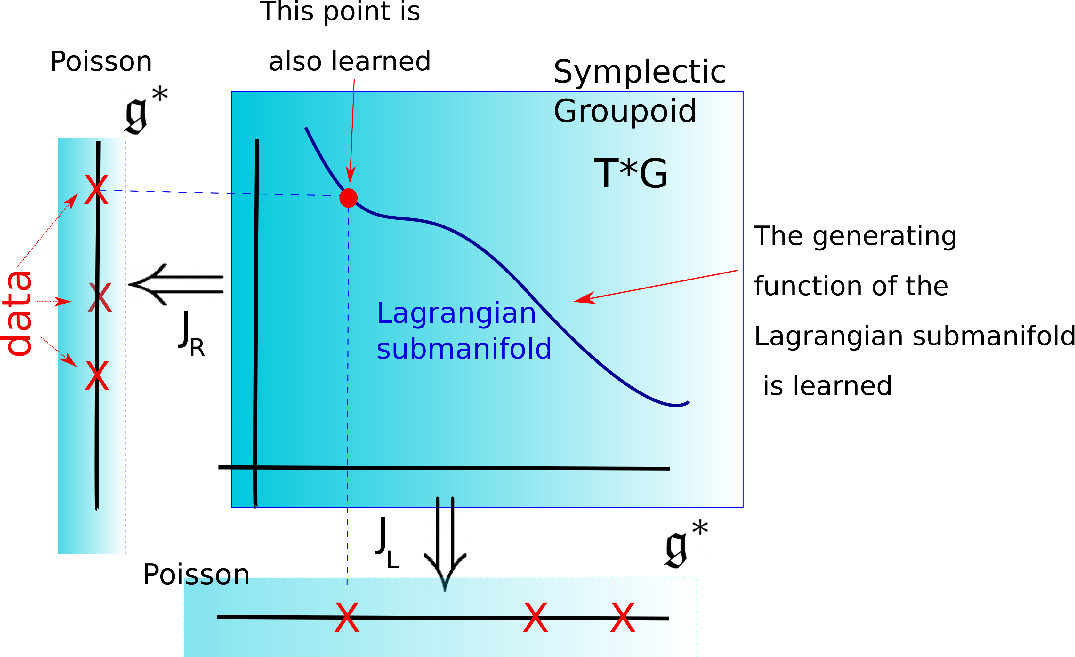}
    \caption{Illustration of the learning process of Poisson manifolds. Both the Lagrangian submanifolds and the point in it connecting two data points are learned at the same time.}
    \label{fig:poss}
\end{figure}

\begin{remark}[Scalability and Alternatives]
    This strategy does not scale well with the number of samples. Thus, when the data set is large another strategy would use another neural network to parameterize the points $p_i$. The exploration of this idea is left for future work.
\end{remark}

\subsubsection{Example: Reduced Rigid Body}
We consider the reduced rigid body in $\mathfrak{so}(3)^*$ and the Hamiltonian given before. The next table shows the MSE achieved in testing data after training a feed-forward neural network of just two layers of $50$ and $10$ neurons.

\begin{table}[H]
\centering
\begin{tabular}{|l|l|l|l|l|}
\hline
 \textbf{Stepsize} & \textbf{N. Samples} & \textbf{MSE} & \textbf{Max. Error} & \textbf{Gradient} \\ \hline \midrule
 \multirow{2}{*}{0.25} & 25 & 0.01297 & 1.69 & 0.0068 \\  \cline{2-5} 
  & 200 & 0.0003 & 0.36 & 0.0018 \\ \hline \midrule
 \multirow{2}{*}{0.5} & 25 & 0.0546 & 1.93 & 0.0058 \\ \cline{2-5}
 & 200 & 0.0006 & 0.3399 & 0.0029 \\ \hline \midrule
 \multirow{2}{*}{1} & 25 & 0.0427 & 1.2597 & 0.0058 \\ \cline{2-5}
  & 200 & 0.0008 & 0.3399 & 0.0029 \\ \hline
\end{tabular}
\caption{Comparison of different scenarios for a neural network trained according to the procedure described here.}
\end{table}

\begin{figure}[H]
    \centering
    \includegraphics[scale = 0.4]{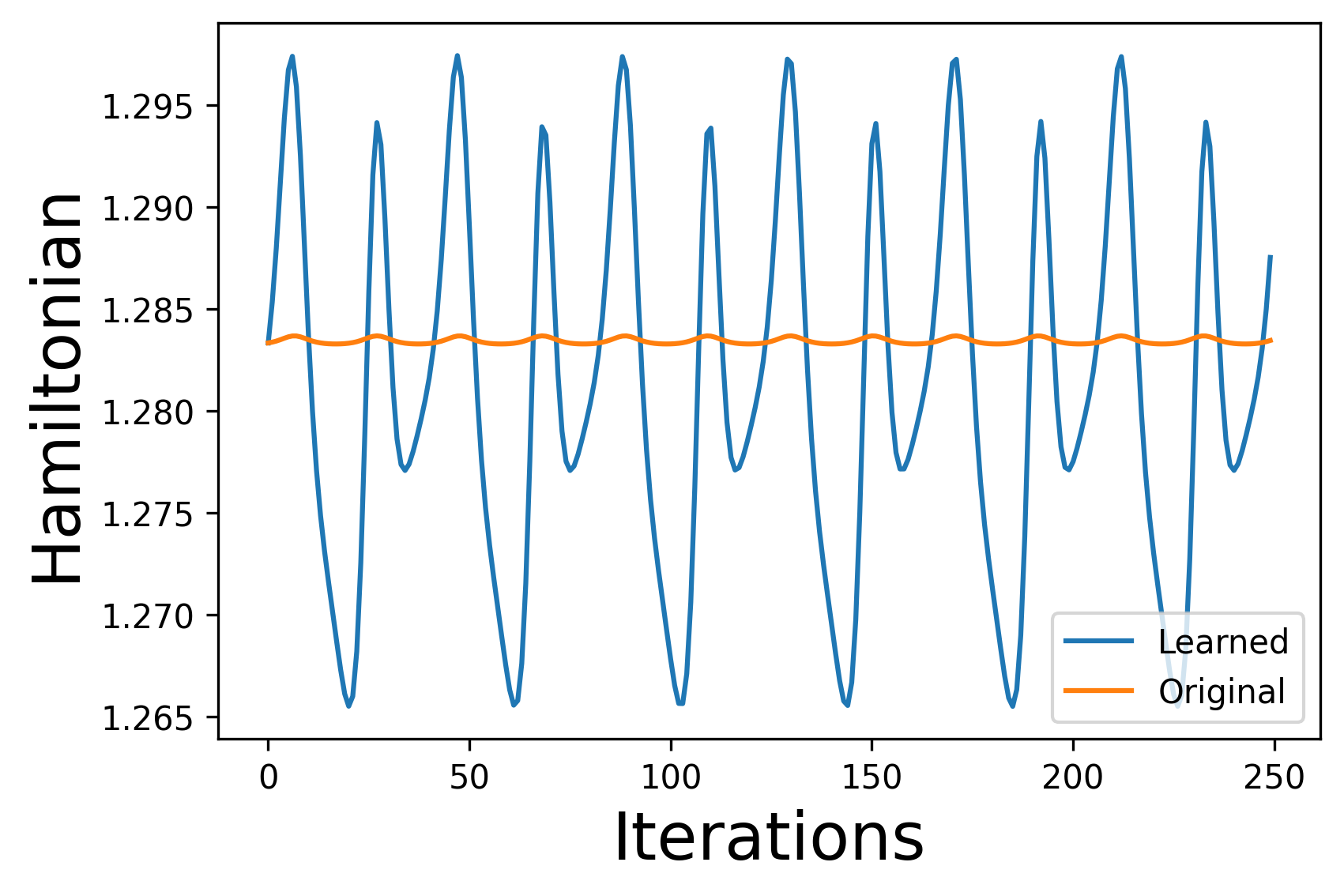}
    \includegraphics[scale = 0.4]{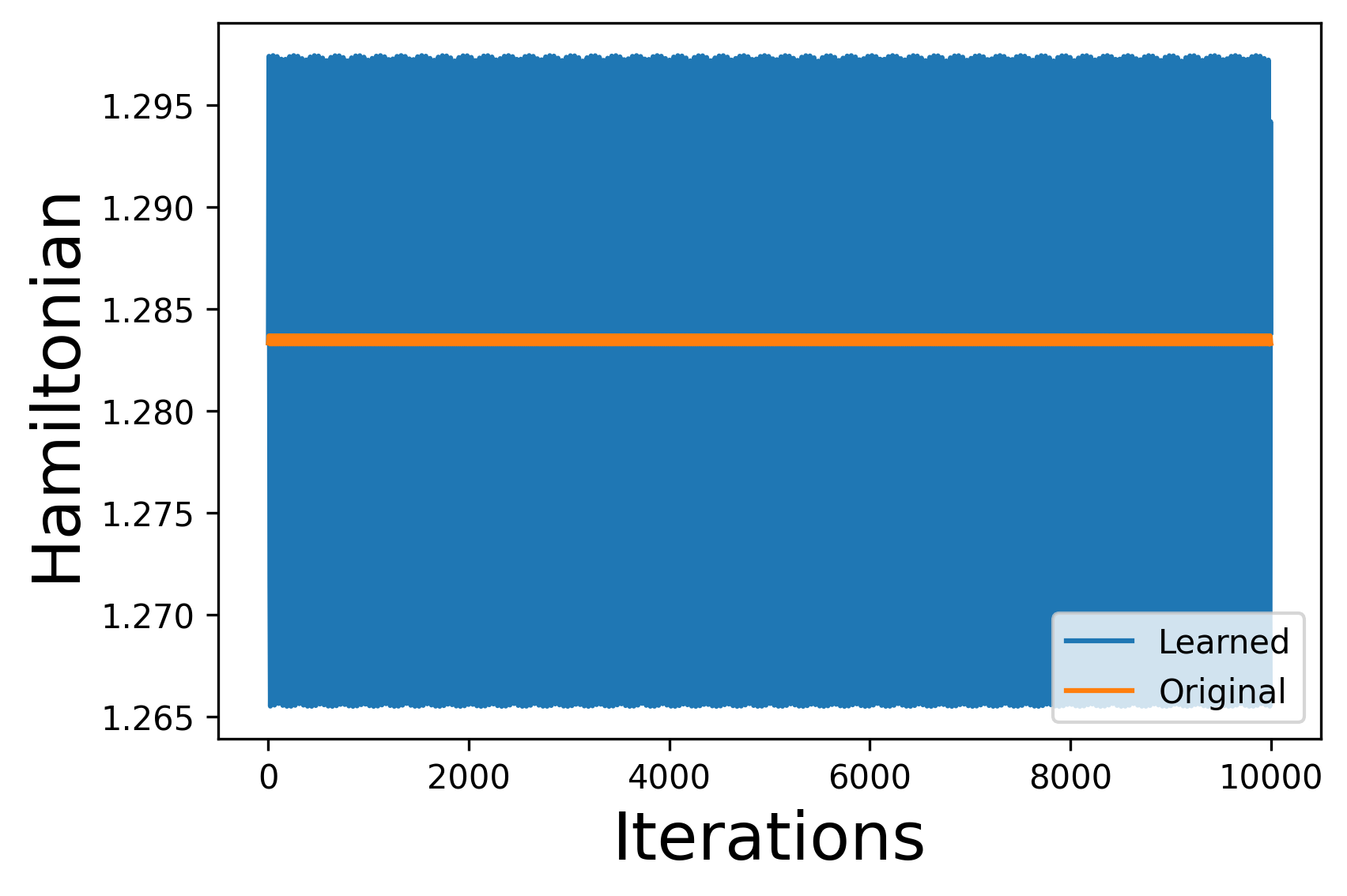}
    \includegraphics[scale = 0.4]{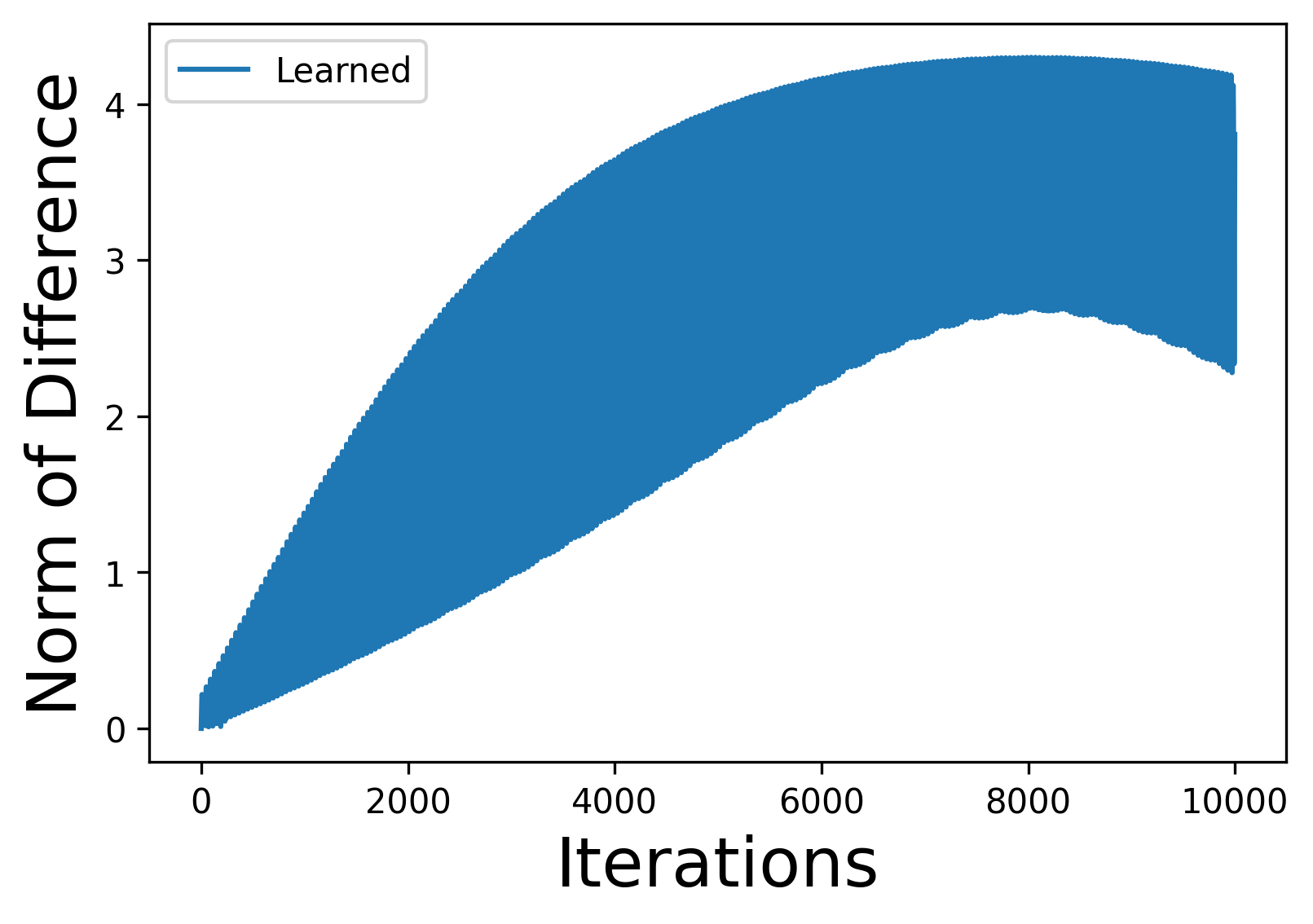}
    \includegraphics[scale = 0.4]{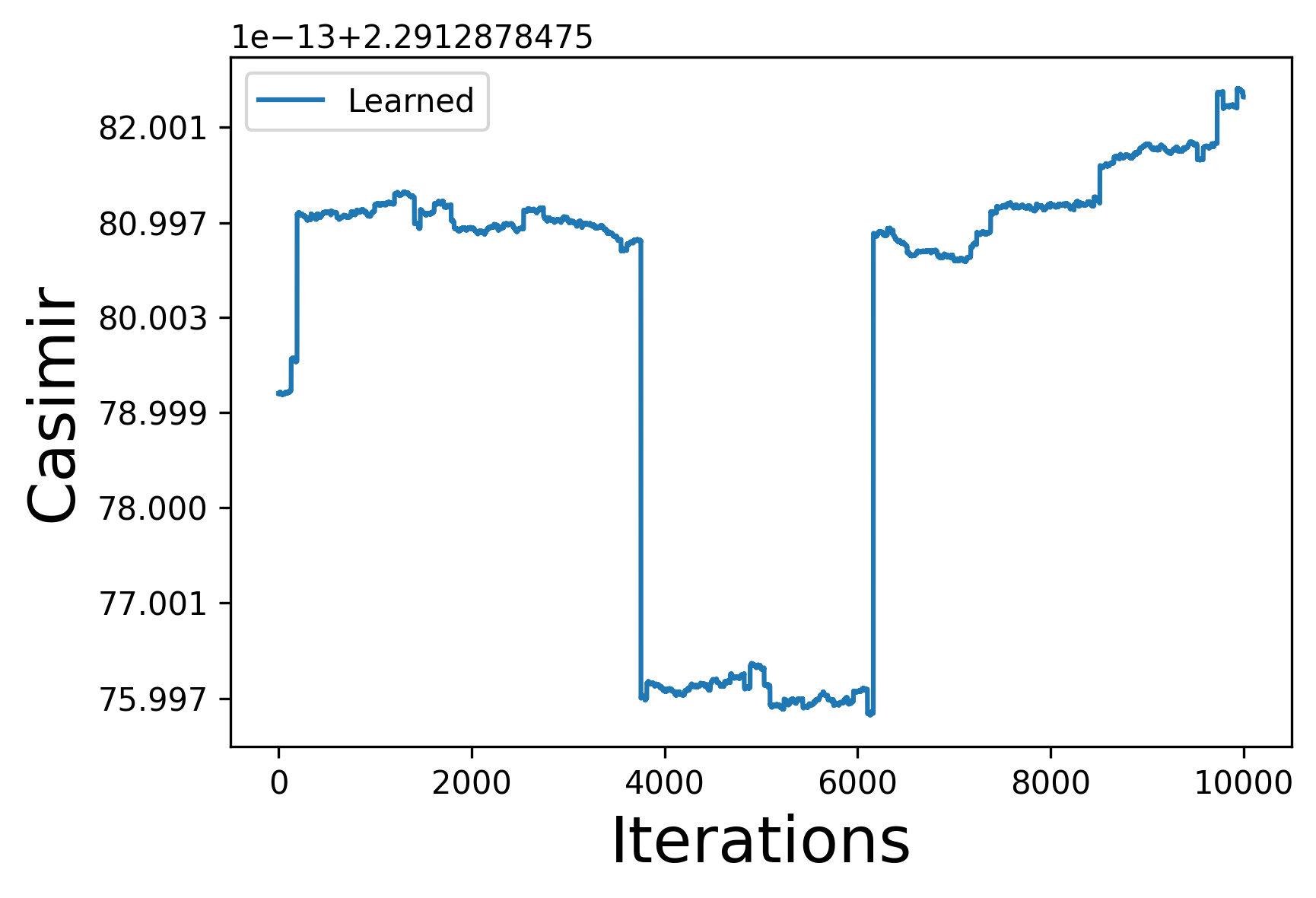}
    \caption{{\it Top-Left:} Comparison of the evolution of the Hamiltonian through a trajectory generated using our symplectic integrators, used for training, and the one predicted using the neural network. We observe how the trained model aims to conserve the Hamiltonian like the original data. {\it Top-Right:} Long term evolution of the Hamiltonian for the original and the trained model. We observe how the behavior depicted in the top-left figure is maintained for long-time simulations. {\it Bottom-Left:} Evolution of the difference in norm for a trajectory between the original model and the trained one. Due to our geometric considerations, our model never escapes the symplectic leaves, as it can be shown due to Casimir conservation in {\it Bottom-Right}. This explains why the difference is bounded even for long-time simulations.
    }
    \label{fig:my_label5}
\end{figure}

Remarkably, the learned dynamics also aim at conserving the energy like the original integrator. The Casimir (norm) is conserved up to rounding error. Our findings are illustrated in Figure~\ref{fig:my_label5}.

\subsection{``Geometrizing'' Integrators}
\begin{figure}[H]
    \centering
    \includegraphics[scale = 0.4]{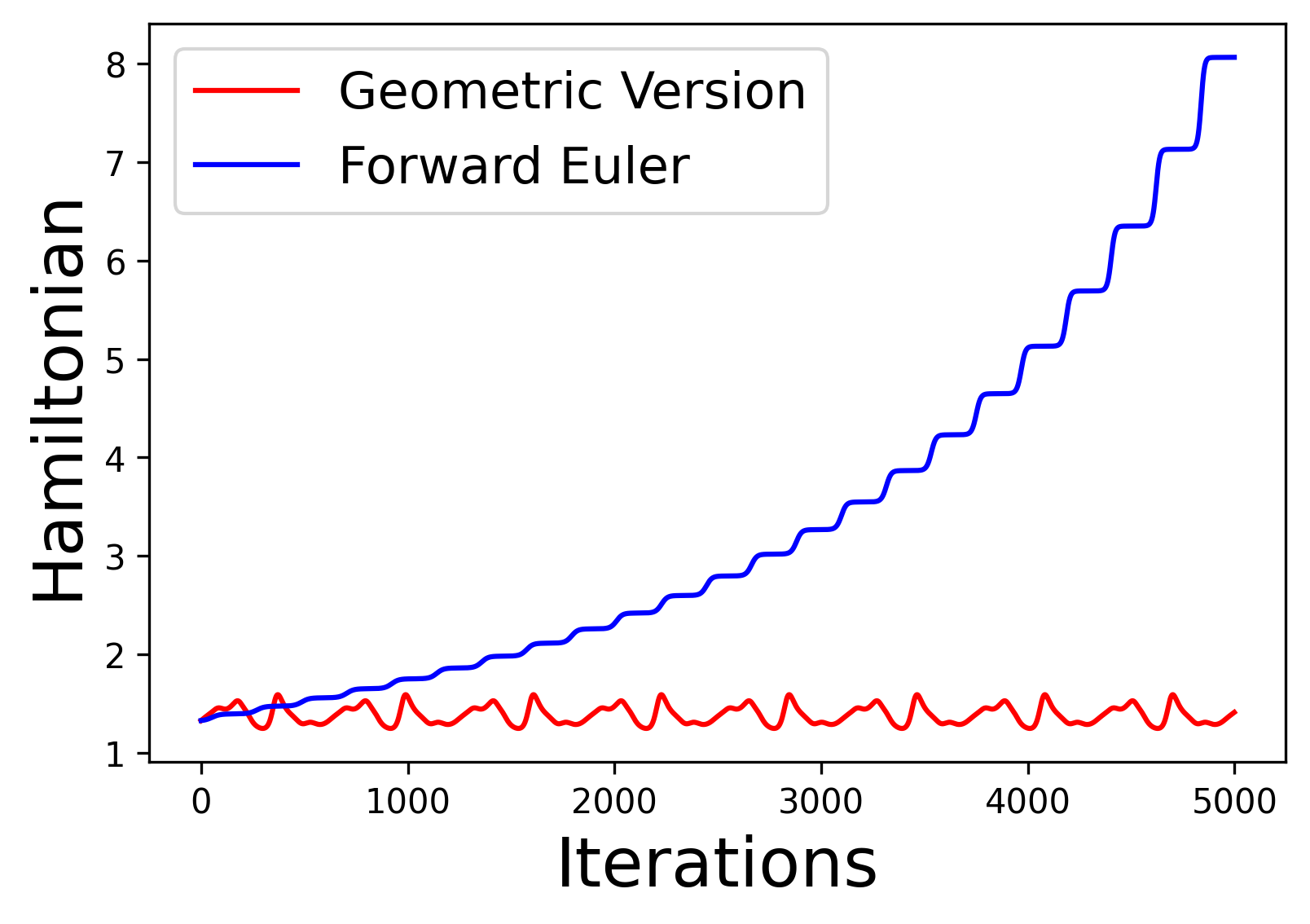}
    \includegraphics[scale = 0.4]{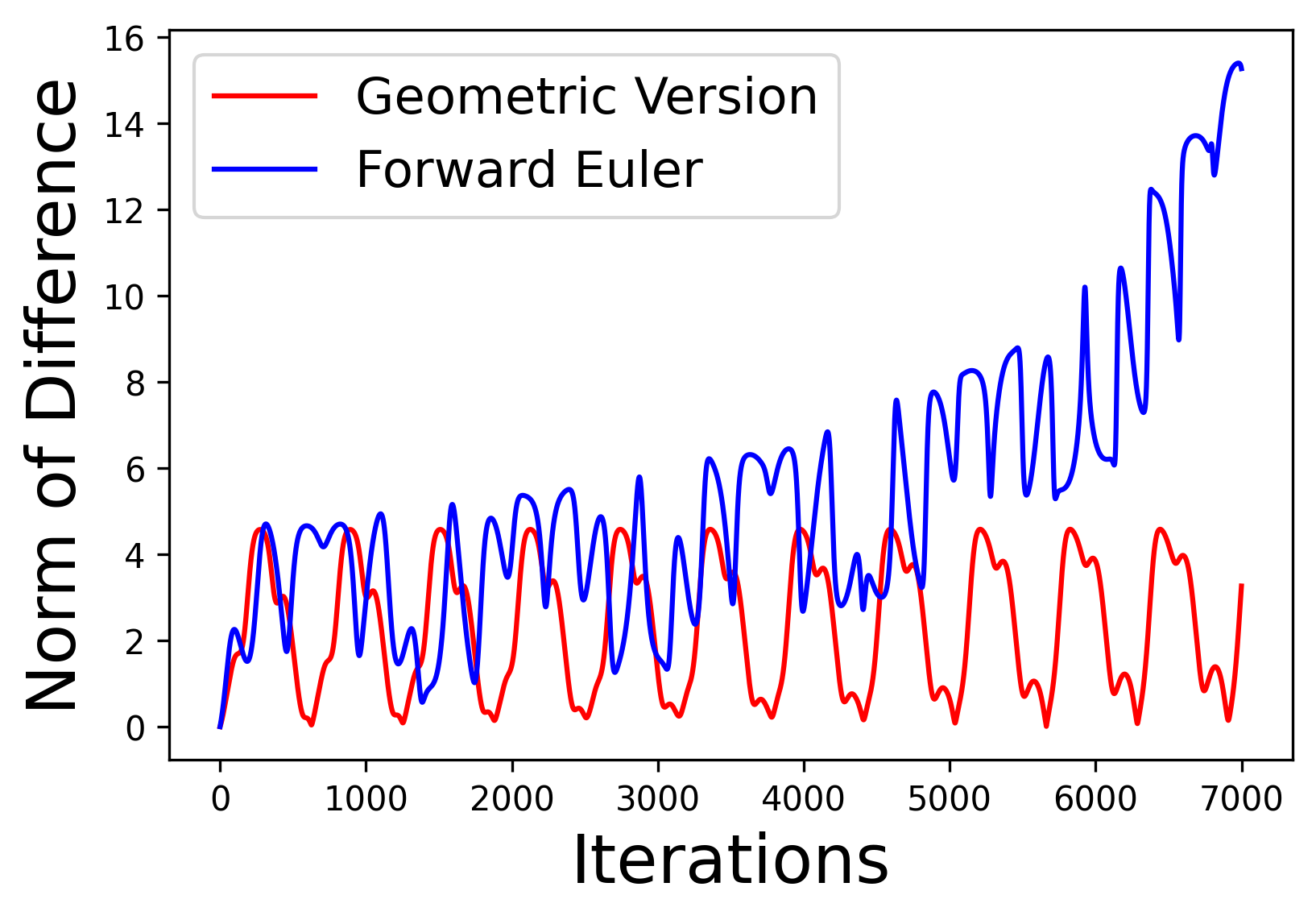}
    \includegraphics[scale = 0.4]{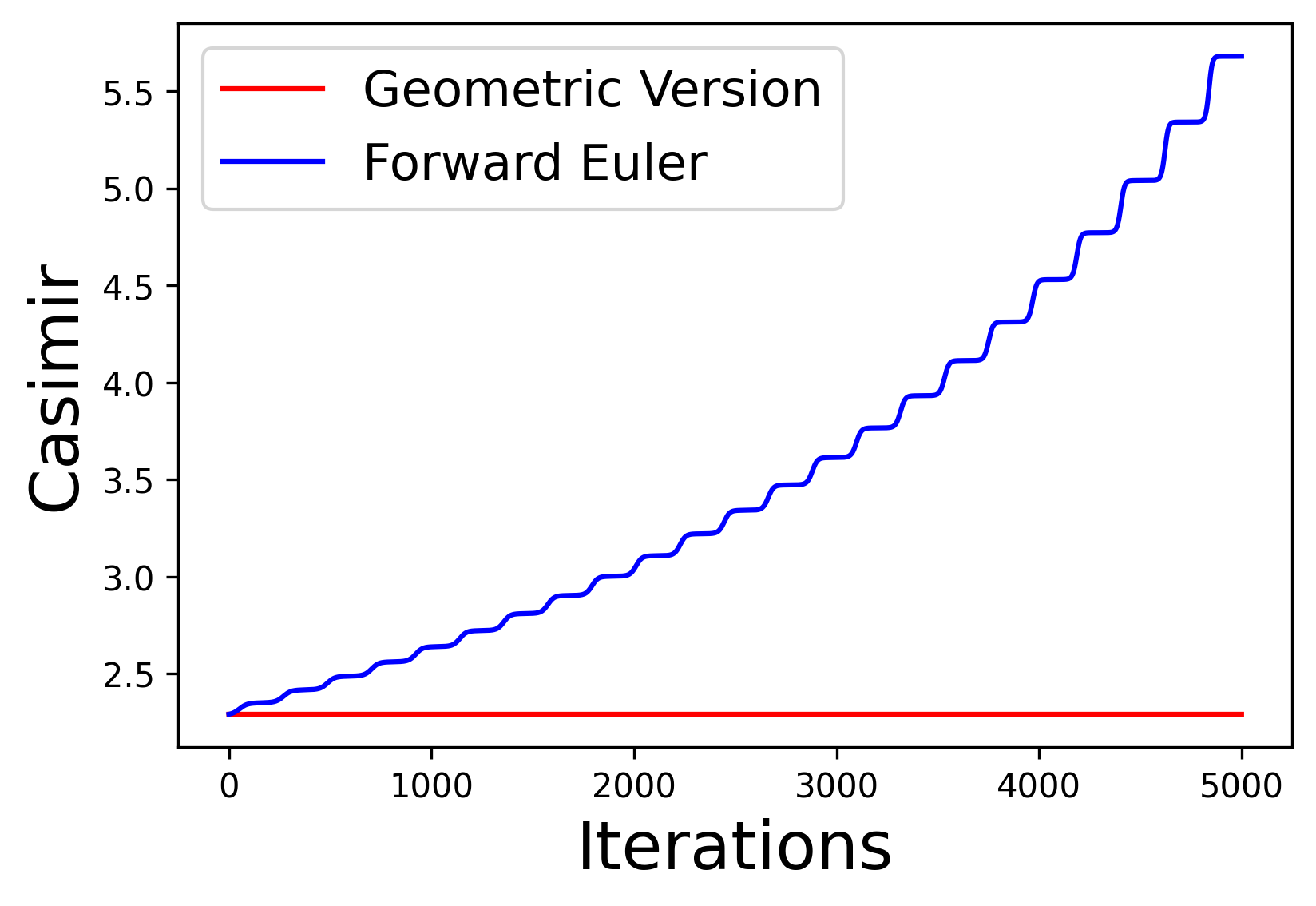}
    \includegraphics[scale = 0.4]{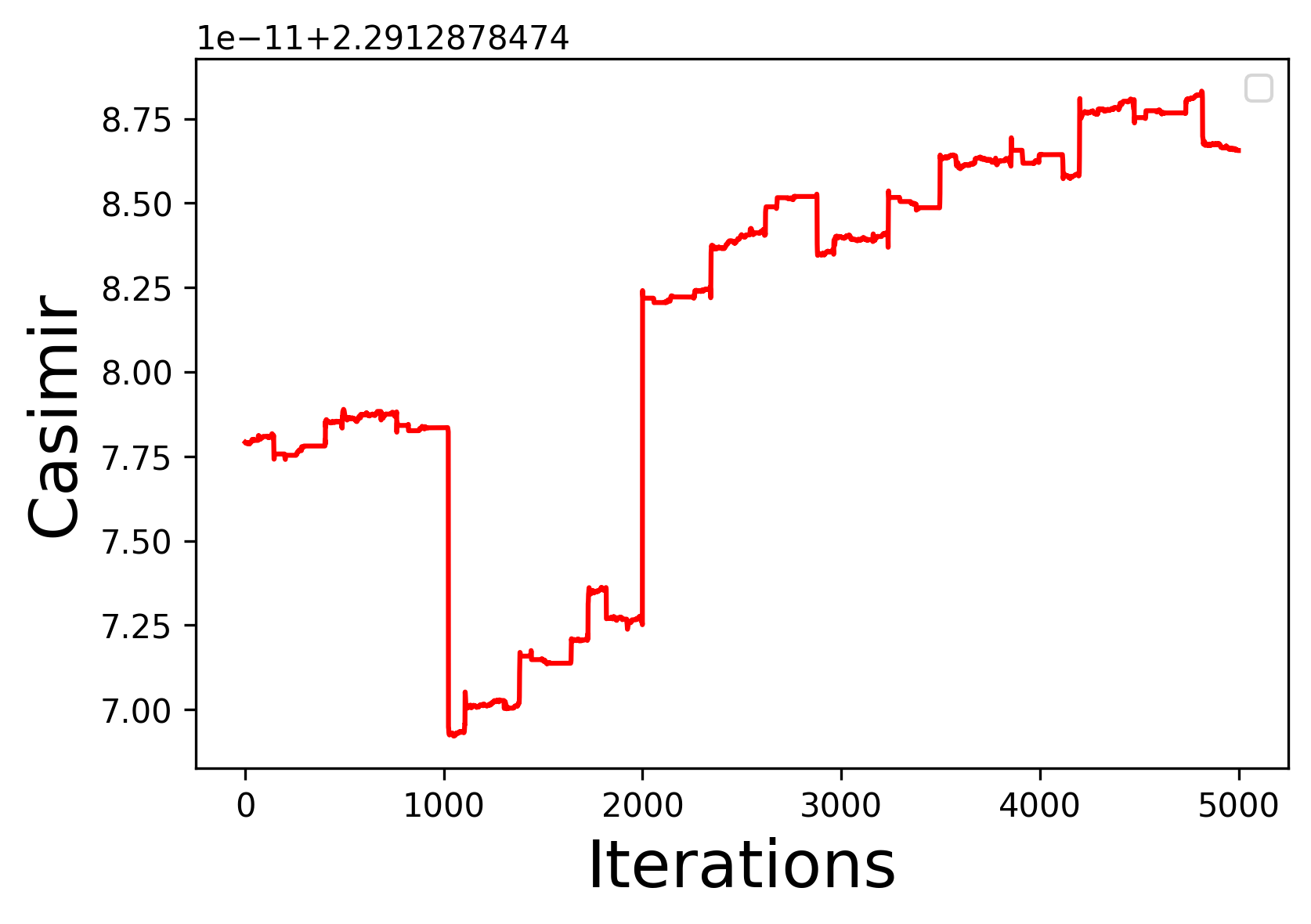}
    \caption{{\it Bottom-Left:} Evolution of the Hamiltonian for the Forward Euler integrator and the trained version. Conspicously, the trained model conserves much better the Hamiltonian, remembering the comparison between geometric and non-geometric integrators. {\it Top-Right:} Evolution of the distance to the real dynamics. We observe how Forward Euler tends to drift after some iterations. The trained model is prevented from getting to far from the real dynamics due to the Casimir conservation shown in {\it Bottom-Left}. Moreove, the conservation of the Casimir is done to round-off error as it can be seen in {\it Bottom-Right}, where we plot the evolution of the Casimir for the trained model for several iterations.}
    \label{fig:my_label6}
\end{figure}

The constructions presented here permit a combination of simulation and learning. Both process can be combined in order to ``geometrize'' integrators that are not geometric or invariant. The idea is to use an integrator to generate a data set and next use the procedures described in this paper to learn geometric versions of them. Although this idea is yet to be fully explored and corresponds to another paper, here we present positive results when applied to simulate Poisson dynamical systems (reduced rigid body). We generated $1000$ pairs of points using Forward Euler integrator to train a feed forward neural network. After that, we take as initial condition the point $(2,0.5,1)$ and compare the evolution of Forward Euler and its geometrized version. The outcome is shown in Figure~\ref{fig:my_label6}

\section{Conclusions and Future Work}\label{sec:Conclusions and Future Work}

In this paper, we have introduced a novel approach for obtaining all $G$-invariant Lagrangian submanifolds by examining the Lagrangian submanifolds in the quotient space. Through our investigations in symplectic and Poisson geometry, we have established a correspondence that can be utilized for simulating and learning Hamiltonian systems. Our findings highlight the potential synergy between differential geometry and machine learning, which is yet to be fully explored. We have demonstrated the efficacy of our results through successful application to the benchmark of the rigid body. Moreover, our constructions lay the foundation for the development of more sophisticated designs to tackle important problems. We present a list of current and future research directions related to our work below.
\begin{itemize}
 \item {\it Control theory.} Geometric control theory has been extensively studied and applied for the last decades. We envision learning paradigms relying on differential geometry able to produce policies through the combination of geometry and reinforcement learning type of techniques.
 \item {\it ``Geometrization'' of integrators.} As far as the authors' knowledge goes, the idea of endowing non-geometric integrators with geometric properties is new. Current research efforts are devoted to the exploration of this idea.
    \item {\it Hamiltonization of non-holonomic dynamical systems.} One of the main points of the present paper is how to approach non-geometric data through geometric structures ($G$-invariant Lagrangian submanifolds). This is in analogy with the phenomenon of Hamiltonization, which can be studied using the techniques presented here.
    
   \item {\it Bayesian framework.} In this study, we utilized neural networks as part of our framework, but it is worth noting that our approach can be integrated with various other settings as well. Recent research has delved into the amalgamation of geometry and Bayesian inference, and there are indications that our results could be extended to incorporate a Bayesian approach to symmetry.

   \item {\it Extension to other Poisson structures.} Our constructions apply Poisson structure of the dual of a Lie algebra. Nonetheless, they rely on the groupoid integrating a Poisson structure, and therefore can be applied in a straightforward way to any dual Lie algebroid. For general Poisson structures, recent advances in (local) integration of Poisson structures should pave the way. We acknowledge that although the exact integration of Poisson structures may be hard, an approximation using series expansions should be enough to obtain numerical integrators and learning paradigms with nice geometric properties.

   \item {\it Construction of Poisson integrators by solving the Hamilton-Jacobi equation using machine learning.} To solve the reduced Hamilton-Jacobi equation 
\[
\displaystyle\frac{\partial S}{\partial t}(t,p) + H(J_R(\frac{\partial S}{\partial p}(t,p),p)) = 0
\]
we can use a Taylor's expansion in the $t$-variable and solve the obtained recurrence, like in~\cite{FLMV}. We are currently exploring the idea of using different approach based on recent advances in PINNs (\cite{pinns_karniadakis}). The main point is to  sample points in the region where we want to solve the equation, obtaining points of the form $(t_i,p_i)$, and then solve the optimization problem in the decision variable $W$
\[
\min\limits_{W}\sum_{i}\left(\displaystyle\frac{\partial S}{\partial t}(t_i,p_i;W) + H(J_R(\frac{\partial S}{\partial p}(t_i,p_i;W),p_i))\right)^2.
\]
This would provide an approximate solution to the Hamilton-Jacobi equation, that can be used the generate a Lagrangian bisection. This Lagrangian bisection induces a Poisson transformation close to the original Poisson flow.
   
\end{itemize}

\section*{Acknowledgements}
The authors acknowledge finantial support from the Spanish Ministry of Science and Innovation under grants PID2022-137909NB-C21, RED2022-134301-TD and the Severo Ochoa Programme for Centres of Excellence in R\&D (CEX2019-000904-S).

\bibliography{References1.bib}
\bibliographystyle{acm}

\end{document}